\documentclass{article}

\usepackage[nonatbib,preprint]{neurips_2025}

\usepackage{url}
\usepackage{amsfonts}       % blackboard math symbols
\usepackage{nicefrac}       % compact symbols for 1/2, etc.
\usepackage{mathtools}      % microtypography
\usepackage{color,bm}
\usepackage{mathrsfs}%huati font
\usepackage{amsmath}%split
\usepackage[linesnumbered,ruled,vlined]{algorithm2e}%algorithm
\usepackage{graphicx} 
\usepackage{multirow}
\usepackage{amssymb}
\usepackage{mathtools}
\usepackage{amsthm}
\usepackage[utf8]{inputenc} % allow utf-8 input
\usepackage[T1]{fontenc}    % use 8-bit T1 fonts
\usepackage{hyperref}       % hyperlinks
\usepackage{url}            % simple URL typesetting
\usepackage{booktabs}       % professional-quality tables
\usepackage{microtype}      % microtypography
\usepackage{xcolor}         % colors
\usepackage[skins]{tcolorbox}

\newcommand{\tx}{\bm{x}}

\newcommand{\tv}{\bm{v}}
\newcommand{\tw}{\textbf{w}}

\newtheorem{definition}{Definition}
\newtheorem{theorem}{Theorem}
\newtheorem{lemma}{Lemma}
\newtheorem{proposition}{Proposition}
\newtheorem{corollary}{Corollary}

\title{On Broken Symmetry in Cognition}

\author{%
  Xin~Li\thanks{This work was partially supported by NSF IIS-2401748 and BCS-2401398. The author has used ChatGPT models to assist in the development of theoretical ideas presented in this paper.} \\
  Department of Computer Science\\
  University at Albany\\
  Albany, NY 12222 \\
  \texttt{xli48@albany.edu} \\
}

\begin{document}

\maketitle

\begin{abstract}
Cognition is not merely the passive accumulation of information but the active resolution of uncertainty through the strategic breaking of symmetry. This paper argues that both cognitive evolution and development proceed through a sequence of symmetry-breaking transitions—disrupting inherent regularities in space, time, self-other boundaries, and representational abstraction. First, bilateral body plans and spatial coding mechanisms, such as grid and place cells, break spatial symmetry by privileging front-facing orientation and localized positional encoding. Second, reinforcement learning introduces temporal asymmetry by prioritizing future outcomes over past experiences, establishing a directed temporal gradient. Third, goal simulation breaks the spatiotemporal symmetry between the internal self-model and the external world, enabling sensorimotor learning and solving Minsky’s search problem via embodied inference. Fourth, mentalizing and imitation learning introduce social asymmetry, allowing agents to model others' beliefs and behaviors, thus breaking the symmetry between minds. Finally, language imposes a recursive, linear structure on unordered thought, breaking the temporal symmetry of expression through syntax and grammar. 
We unify these forms of asymmetry under the \emph{Context-Content Uncertainty Principle (CCUP)}, which frames cognition as a cyclic process that minimizes conditional entropy by transforming high-entropy contexts into low-entropy, goal-relevant content. Central to this process is the principle of \emph{structure-before-specificity}, a cognitive strategy that first encodes ambiguous input into stable latent structure before binding it to specific observations. This strategy enables generalization, reduces sample complexity, and avoids overfitting. Moreover, by inverting inference (i.e., reconstructing latent context from observed content), cognition breaks the curse of dimensionality through structured representations that scaffold efficient planning and abstraction. In this view, symmetry breaking is not incidental but the foundational mechanism by which cognition organizes, stabilizes, and scales inference in an uncertain and dynamic world.
\end{abstract}

%\vspace{-0.2in}
\section{Motivation: Symmetry Breaking as the Engine of Cognitive Evolution}
%\vspace{-0.1in}
Symmetry is a foundational concept in both natural and cognitive sciences. In physics, symmetry breaking underlies the emergence of structure in the universe, from the differentiation of forces in the early cosmos to the formation of complex systems \cite{anderson1972more}. In biology, symmetry breaking gives rise to organismal form, function, and behavior, enabling specialization and adaptation \cite{gierer1972theory}. In cognitive science, symmetry breaking refers to the emergence of directional, lateralized, and hierarchical patterns in neural structure and function that allow the brain to specialize, differentiate, and adapt to environmental demands \cite{ocklenburg2024lateralized}. We argue that cognition itself is fundamentally organized through the progressive breaking of symmetry, a principle that governs not just the structure of the brain but the flow of information, the formation of representations, and the dynamics of inference.

Evolution does not design with foresight \cite{dawkins1996blind}; instead, it reuses what works. As Buzsáki notes, ``if a constructive mechanism is invented by nature in simple organisms, more complex animals tend to exploit it'' (pp. 259, \cite{buzsaki2006rhythms}). This process, known as bootstrapping, is a hallmark of cognitive evolution \cite{bennett2023brief}, which can be understood as a sequence of symmetry-breaking events that progressively reduce uncertainty, constrain inference, and expand representational capacity. Each major breakthrough in cognition, from spatial navigation to social communication, emerged not through a linear accumulation of complexity but through a fundamental shift in representational asymmetry. These symmetry breaks transformed initially undifferentiated cognitive processes into structured, directional systems capable of inference, generalization, and communication. 

We propose that the development and evolution of cognitive function can be understood as a systematic sequence of symmetry-breaking operations across multiple domains: space, time, self-world model, social boundaries/structures, and language \cite{bennett2023brief}. Each break resolves specific bottlenecks in learning, prediction, or communication by introducing asymmetries that reduce uncertainty and guide inference \cite{buzsaki2006rhythms}. From the privileging of forward-facing spatial orientation to the recursive structuring of language, these symmetry breaks encode constraints that transform high-entropy, ambiguous contexts into low-entropy, goal-directed content \cite{gyorgy2019brain}.
Following the five breakthroughs given in \cite{bennett2023brief}, we have 

\textbf{1) Steering: Spatial Symmetry Breaking.}
The evolution of bilateral body plans and forward-facing sensorimotor architectures broke spatial symmetry by privileging the front over the back. By prioritizing the front over the back, organisms gained a directional bias that allowed them to reduce spatial uncertainty and better align perception with action \cite{burgess2008spatial}. This spatial asymmetry was soon mirrored in time: as brains evolved, they began to privilege forward associations over backward ones, paving the way to temporal asymmetry. 

\textbf{2) Reinforcing: Temporal Symmetry Breaking.}
Reinforcement learning introduced an asymmetry in time by prioritizing future rewards based on past experiences. This broke the symmetry between past and future, establishing a predictive gradient that guided decision-making and planning. Temporal discounting compressed the representational space of possible futures, allowing cognitive systems to focus on near-term outcomes \cite{watkins1992q}. %Under CCUP, reinforcement learning minimized uncertainty by aligning historical context (\( \Psi \)) with goal-conditioned content (\( \Phi \)), forming a temporal arrow of inference.

\textbf{3) Simulating: Spatiotemporal Symmetry Breaking.}
Goal simulation extended symmetry breaking into latent space by separating the self-model from the world-model. Rather than inferring passively from external stimuli, cognitive systems began simulating hypothetical futures and back-inferencing the conditions necessary to achieve them. This broke the symmetry of direct perception by enabling internal counterfactual reasoning \cite{pearl2009causality}. The emergence of dorsal-ventral separation in cortical streams reflects this architectural asymmetry, allowing the brain to disambiguate external context while internally generating goal-driven content.

\textbf{4) Mentalizing: Social Symmetry Breaking.}
The ability to model other agents' internal states broke the symmetry between self and others. Through mentalizing and imitation, cognitive systems began treating other individuals not as behaviorally equivalent, but as sources of latent goals, beliefs, and intentions. This representational asymmetry supported recursive inference, cultural learning, and the emergence of Theory of Mind (ToM) \cite{leslie2004core}. %Under CCUP, mentalizing reduces uncertainty by aligning observed social context (\( \Psi \)) with inferred internal models of others’ content (\( \Phi \)), enabling stable coordination and communication.

\textbf{5) Speaking: Linguistic Symmetry Breaking.}
Language imposed a linear, compositional structure onto unordered mental representations, breaking the symmetry of thought and enabling externalization. Syntax and grammar constrain both the generation and interpretation of meaning, allowing ambiguous mental content to be compressed into structured linguistic context, and vice versa \cite{levinson1983pragmatics}. This broke the symmetry between internal simulation and external communication, allowing cognition to scale socially and culturally. %Language completes the CCUP inference hierarchy across agents by stabilizing shared content and minimizing mutual uncertainty.

%\paragraph{Summary.}
%Each cognitive advance—steering, reinforcing, simulating, mentalizing, and speaking—represents a symmetry break that transformed cognition by introducing structure, directionality, and constraints. These breaks enabled the dynamic alignment of context and content across space, time, agency, social structure, and symbolic systems. Viewed through the lens of CCUP, symmetry breaking is not a side effect of evolution—it is the \emph{engine} that drives cognitive inference, complexity, and communication. It reduces entropy not by adding information, but by \emph{organizing} it—bootstrapping the brain into a model-building, goal-seeking, and meaning-sharing system

In this paper, we unify cognitive asymmetries under the \emph{Context-Content Uncertainty Principle (CCUP)} \cite{li2025CCUP}, which models cognition as a cyclic process that transforms high-entropy contexts into low-entropy, goal-relevant content. At its core is the principle of \emph{structure-before-specificity} \cite{clark2013whatever}: ambiguous input is first compressed into a stable structure before being bound to specific observations. This strategy supports generalization, lowers sample complexity, and enables scalable inference. Through inverted inference \cite{friston2017active}, reconstructing context from content, cognition breaks the curse of dimensionality by leveraging structured representations for efficient planning and abstraction. Symmetry breaking, in this view, is the engine that stabilizes and scales cognitive inference.
%we show how these symmetry-breaking events, spanning spatial, temporal, spatiotemporal, social, and linguistic domains, emerge as necessary adaptations under the Context-Content Uncertainty Principle (CCUP) framework \cite{li2025CCUP}. Each break transforms high-entropy contexts into structured, low-entropy content by selectively constraining inference. %Together, they form a complementary learning system \cite{mcclelland1995there}, explaining both the speed and stability of human cognition, paralleling what has been framed as fast and slow thinking in cognitive science \cite{kahneman2011thinking}.

%\vspace{-0.1in}
\section{The Context-Content Uncertainty Principle (CCUP)} 
%\vspace{-0.1in}
\label{sec:2}

At the heart of cognition lies the task of interpreting and acting upon ambiguous sensory and internal signals under uncertainty. A central challenge is how to form stable, goal-relevant representations (\emph{content}) from high-entropic, variable input (\emph{context}). In many cognitive domains, ranging from perception and memory to reasoning and language, this task is constrained by the limited bandwidth of sensors, motors, and memory systems. These limitations necessitate selective information processing, prioritizing some variables (e.g., task-relevant or goal-specific) over others \cite{niv2019learning}.
The \textbf{Context-Content Uncertainty Principle (CCUP)} formalizes this process by characterizing cognition as a dynamic alignment between \emph{context} (high-entropy, ambiguous conditions, denoted \( \Psi \)) and \emph{content} (low-entropy, selected representations, denoted \( \Phi \)) \cite{li2025CCUP}. Importantly, this alignment is \emph{not symmetric}: context often occupies a compressed sensory or environmental space but remains ambiguous due to aliasing and interference, whereas content spans a semantically rich latent space but is shaped by goal-driven constraints that stabilize its entropy.

%\paragraph{Formal Statement and Theorem.}
Let \( \Psi \) denote context variables (e.g., ambiguous sensory input, environmental cues), and \( \Phi \) denote content variables (e.g., latent causes, goal states, task-relevant representations). CCUP asserts a fundamental trade-off in inference: the system seeks to minimize both 1) \textbf{specification/disambiguation}: conditional entropies \( H(\Phi \mid \Psi) \) (inference/encoding); and 2) \textbf{generalization/reconstruction}: \( H(\Psi \mid \Phi) \)  (generation/decoding). The main result of CCUP is stated as follows:

\begin{lemma}[CCUP Lower Bound]
Let \( \Psi \) and \( \Phi \) be random variables representing context and content. Then, we have
$H(\Phi \mid \Psi) + H(\Psi \mid \Phi) \geq H(\Psi, \Phi) - I(\Psi; \Phi)$
with equality if and only if the joint distribution \( p(\Psi, \Phi) \) is cycle-consistent and bidirectionally minimal.
\end{lemma}

\textbf{Remark:} \( I(\Psi; \Phi) \) is the mutual information between context and content and \( H(\Psi, \Phi) \) is their joint entropy. The inequality highlights the irreducible uncertainty in attempting to map between context and content unless a consistent inference cycle is established.
%\paragraph{Interpretation and Cognitive Implications.}
The CCUP lower bound provides a formal backbone for understanding why cognition must \emph{break symmetry}. In real-world conditions, the bidirectional mappings between \( \Psi \) and \( \Phi \) are lossy: perception suffers from aliasing \cite{wehner1987matched}, memory from interference \cite{mccloskey1989catastrophic}, and action from feedback ambiguity \cite{todorov2002optimal}. These conditions make it impossible to achieve a perfect mutual mapping, thereby necessitating structural solutions such as \emph{asymmetric inference cycles} to iteratively minimize uncertainty.

\paragraph{From CCUP to the Strategy of Generalization-before-Specification.}
The CCUP reveals a fundamental asymmetry in cognition: context variables \( \Psi \) are high-entropy and ambiguous, while content variables \( \Phi \) are lower-entropy and selectively encoded. This asymmetry implies that intelligent systems cannot infer specific content directly from raw context without incurring high uncertainty or representational interference. Instead, the system must first compress across multiple contexts to construct a generalizable structure that captures latent invariants \cite{bengio2013representation}. Only then can it safely bind specific experiences to that structure. This two-stage process, first minimizing \( H(\Psi|\Phi) \) to extract structure, then minimizing \( H(\Phi| \Psi) \) for context-sensitive binding, forms a robust strategy for reducing uncertainty across asymmetric inference cycles. It reflects a core principle of learning and memory in intelligent systems: \textbf{structure precedes specificity} or \textbf{generalization-before-specification} \cite{clark2013whatever}.

\begin{theorem}[Structure Precedes Specificity]
Let \( \Psi \) denote high-entropy context variables and \( \Phi \) denote low-entropy content variables inferred under CCUP. Let \( \mathcal{D} = \{ \Psi_i \}_{i=1}^N \) be a set of observed contexts drawn from an environment. Then an optimal inference strategy proceeds in two stages:
1) {Structure Formation (Generalization)}:
    The agent seeks a latent structure \( \Phi \) that minimizes the expected conditional entropy across contexts:
    $\Phi^\ast = \arg\min_{\Phi} \mathbb{E}_{i} \left[ H(\Psi_i \mid \Phi) \right]$, which yields a generalizable representation \( \Phi^\ast \) that captures invariant structure across variable observations;
2) {Experience Binding (Specification)}:
    Once \( \Phi^\ast \) is established, the agent performs context-specific inference by minimizing:
    $H(\Phi^\ast \mid \Psi_j)$ for each new observation \( \Psi_j \in \mathcal{D}_{\text{new}} \), effectively binding content to context using the learned structural prior.
    \label{thm:structure_precedes_specificity}
\end{theorem}

\textbf{Remark:} The proof of the above theorem is referred to in Appendix A. This structure-before-specificity principle implies a fundamental temporal strategy: by delaying the binding of specific observations until a stable latent structure has been formed, cognition enhances its capacity for generalization. In other words, delay before specificity is not a weakness but a necessary condition for robust inference under uncertainty \cite{todorov2002optimal}. Formally, we have

\begin{theorem}[Delay Before Specificity Enables Robust Generalization]
Let \( \Phi^\ast \) be the latent structure learned by minimizing \( \mathbb{E}_i[H(\Psi_i \mid \Phi)] \) as in Theorem \ref{thm:structure_precedes_specificity}. Under CCUP, binding specific experiences prematurely, i.e., before structure formation is complete, leads to poor generalization and representational interference.
Formally, let \( \Phi_{\text{early}} \) be a representation inferred before cross-context compression is complete, \( \Phi_{\text{late}} = \Phi^\ast \) be a structure inferred after minimizing entropy across diverse \( \Psi_i \).
For a new context \( \Psi_j \), the bound on inference error satisfies:
$\mathbb{E} \left[ H(\Phi_{\text{early}} \mid \Psi_j) \right] > \mathbb{E} \left[ H(\Phi^\ast \mid \Psi_j) \right]$.
\label{cor:delay_enables_generalization}
\end{theorem}

\textbf{Remark:} The proof of the above theorem is referred to Appendix B. Without delaying binding until structure formation stabilizes, the agent suffers from the risks of overfitting, high content-context misalignment, and reduced transfer to novel contexts. Under CCUP, delaying specificity until a generalizable structure \( \Phi^\ast \) is learned is essential for minimizing inference error and enabling robust generalization. This computational delay is the root of inverted/active inference \cite{friston2017active}, reflecting the biological principle of separating memory encoding and consolidation phases, such as in the sleep-wake cycle \cite{hinton1995wake} or neocortical-hippocampal interaction \cite{buzsaki1996hippocampo}.
%In cognitive science, this principle offers a unifying explanation for a wide range of empirical phenomena: 1) \textbf{Perception}: Sensory systems stabilize object identity (content \( \Phi \)) despite ambiguous scenes (context \( \Psi \)), via bottom-up encoding and top-down decoding; 2) \textbf{Memory}: Episodic memory encodes contextual information (\( \Psi \)), while semantic memory distills stable content (\( \Phi \)) over time, minimizing \( H(\Psi \mid \Phi) \) through consolidation; 3) \textbf{Sensorimotor learning}: Contextual feedback is aligned with motor goals through perception-action cycles that iteratively reduce uncertainty between \( \Psi \) and \( \Phi \); 4) \textbf{Mentalizing}: Social contexts are inferred from minimal cues by simulating internal content states (beliefs/intentions), requiring cyclical ToM modeling between \( \Psi \) and \( \Phi \); 5) \textbf{Language}: Syntax and grammar serve to minimize ambiguity in mapping structured linguistic sequences (content \( \Phi \)) to speaker intent and listener interpretation (context \( \Psi \)). Each of these cases reflects an informational asymmetry that is resolved through symmetry breaking, spatially, temporally, socially, or representational, in a manner consistent with CCUP.

\begin{tcolorbox}[colback=gray!5!white, colframe=blue!75!white, title=Uncertainty Principle: Context vs. Content Representation]
The CCUP provides a foundational framework for understanding cognitive systems by highlighting a fundamental informational asymmetry between context and content: context refers to ambiguous, high-entropy environmental or internal conditions, while content denotes specific, low-entropy, goal-relevant representations. CCUP asserts that cognition systematically resolves this asymmetry by iteratively reducing conditional uncertainties, transforming context into structured content via cycle formation and inverted/active inference. The computational strategy of \textbf{structure-before-specifity} implies an \textbf{asymmetrical and inverted} inference cycle: first minimizing \( H(\Psi \mid \Phi) \) to extract structure (generalization/decoding) and then minimizing \( H(\Phi \mid \Psi) \) for context-sensitive binding (specification/encoding).
\end{tcolorbox} 

\section{Spatial Symmetry Breaking: the Origin of Goal-Directed Behavior}
\label{sec:3}

\paragraph{Structure Must Precede Specificity.} 
At the heart of the CCUP lies the idea that cognition proceeds by minimizing uncertainty through the iterative alignment of high-entropy \emph{context} variables \( \Psi \) with low-entropy \emph{content} variables \( \Phi \), which implies a foundational computational strategy in spatial cognition:
\textbf{first construct a generalizable structure (grid), then bind it to specific experiences (places).}
%\paragraph{Grid Cells Provide Low-Entropy Content}.
Grid cells generate a periodic, translation-invariant code over continuous space \cite{moser2008place}. While ambiguous in isolation due to their periodicity, this code possesses low entropy relative to the space of all sensory trajectories and supports metric inference over spatial displacements.
From the CCUP perspective, grid cells provide a \emph{prior over content}:
$H(\Phi_{\text{grid}}) \ll H(\Psi)$.
%This structural prior constrains the space of plausible spatial representations, enabling efficient inference.
%\subsection*{Place Cells Bind Grid Structure to Disambiguated Contexts}
Place cells integrate the structured periodic code \( \Phi_{\text{grid}} \) from grid cells with disambiguating sensory and directional context \( \Psi \) \cite{moser2015place}. This binding operation breaks directional symmetry by localizing representations into place fields and reduces the conditional entropy:
$H(\Phi_{\text{place}} \mid \Psi) \downarrow$.    
Thus, place-cell specificity emerges \emph{only after} a structured content space has been established, permitting robust generalization.

\paragraph{Asymmetric Inference Cycle: General Structure Enables Specific Inference.}
The inference process proceeds as a cycle:
1) grid cells propose a structured content hypothesis \( \Phi_{\text{grid}} \); 2) place cells specialize this code by conditioning on context \( \Psi \); 3) a cycle of refinement aligns \( \Psi \) and \( \Phi \) through bidirectional prediction: $\Phi_{\text{grid}} \rightarrow \Phi_{\text{place}} \rightarrow \Psi \rightarrow \Phi_{\text{grid}}$.
This inference loop enables generalization across environments, rapid learning from sparse experience, and stabilization of spatial memory under contextual ambiguity \cite{whittington2022build}. Grid cells instantiate a low-entropy, structured content space that is essential for the emergence of high-resolution, context-specific place fields. Without this structured prior, inference from high-entropy, aliased context would be underdetermined. Formally, we have

\begin{proposition}
[Bootstrapping Order of Symmetry Breaking in Spatial Cognition]
Let \( \mathcal{X} \subset \mathbb{R}^2 \) denote continuous 2D space, and  \( \Phi_{\text{grid}} \)/\( \Phi_{\text{place}} \) denote content representations encoded by grid cells and place cells, respectively. Let \( \Psi \) denote high-entropy contextual inputs (e.g., heading direction). The emergence of structured spatial inference requires the following bootstrapping order of symmetry breaking:
1) {Translational symmetry breaking:} grid cells first break the translational symmetry of \( \mathcal{X} \) by imposing a periodic, low-entropy metric code:
    $\Phi_{\text{grid}}: \mathcal{X} \rightarrow \mathbb{T}^n$,
    where \( \mathbb{T}^n \) is an $n$-dimensional toroidal lattice representing phase codes over space.
2) {Directional symmetry breaking:} place cells subsequently break the symmetry of heading by binding grid codes to disambiguated context \( \Psi \), forming localized fields:
    $\Phi_{\text{place}} = f(\Phi_{\text{grid}}, \Psi), H(\Phi_{\text{place}} | \Psi) \downarrow$;
3) {Inference bootstrapping:} This hierarchical symmetry breaking minimizes the total uncertainty across context-content cycles:
   $ H(\Phi_{\text{place}} | \Psi) + H(\Psi | \Phi_{\text{place}}) > H(\Phi_{\text{grid}} | \Psi) + H(\Psi | \Phi_{\text{grid}})
    \Rightarrow \Phi_{\text{grid}} \text{ precedes } \Phi_{\text{place}}$.
\end{proposition}

\textbf{Remark:} Grid-cell symmetry breaking is a computational and evolutionary prerequisite for the emergence of place-cell representations \cite{solstad2006grid}. This bootstrapping order reflects nature's strategy of reducing spatial entropy through periodic encoding before enabling localized, context-specific memory representations. Next, we take turns to study translational and directional symmetry breaking.

\paragraph{Translational Symmetry Breaking by Grid Cells.} 
Grid cells define a low-entropy scaffold in an otherwise uniform space \cite{bush2015using}: 1) raw physical space is translationally symmetric since there are no intrinsic features that distinguish one point from another; 2) translational symmetry makes inference ill-posed: without structure, spatial representation becomes degenerate and ambiguous. Nature needs to break translational symmetry first because it imposes the minimal structure required for learning and localization. This structure provides the scaffold on which more complex, context-specific episodic maps (place cells) can be formed.

\begin{proposition}[Grid Cells as Translational Symmetry Breaking in Space]
Let physical space \( \mathcal{X} \subset \mathbb{R}^2 \) be translationally symmetric, i.e., for any displacement vector \( \vec{d} \), locations \( \vec{x} \) and \( \vec{x} + \vec{d} \) are indistinguishable under uniform priors.
Grid cells break this translational symmetry by defining a periodic content code \( \Phi_{\text{grid}}(\vec{x}) \) over \( \mathcal{X} \) such that:
$\Phi_{\text{grid}}(\vec{x}) = \Phi_{\text{grid}}(\vec{x} + n_1 \vec{v}_1 + n_2 \vec{v}_2), \quad \forall n_1, n_2 \in \mathbb{Z}$,
where \( \vec{v}_1, \vec{v}_2 \) are non-parallel basis vectors defining a hexagonal lattice.
\end{proposition}

\textbf{Remark:}
This periodic tiling imposes a low-entropy, spatially structured prior that breaks translational invariance by establishing a discrete internal metric on continuous space. It supports path integration \cite{mcnaughton2006path} by encoding displacement through phase shifts across modules by minimizing conditional entropy \( H(\Psi_{\text{place}} \mid \Phi_{\text{grid}}) \), allowing ambiguous place-cell contexts to be inferred from structured grid-cell content \cite{solstad2006grid}.
This way, grid cells implement a functional symmetry-breaking that transforms a homogeneous spatial manifold into a computable lattice, enabling error correction \cite{sreenivasan2011grid} by a multiscale specification as follows.

\begin{corollary}[Multi-Scale Grid Modules for Redundant Disambiguation]
Let \( \{ \Phi^{(i)}_{\text{grid}} \}_{i=1}^M \) denote a set of grid-cell modules, each encoding spatial position \( \vec{x} \in \mathbb{R}^2 \) with periodicities \( \lambda_i \), where \( \lambda_1 < \lambda_2 < \dots < \lambda_M \).
Each module independently tiles space with its own hexagonal lattice, introducing phase ambiguity:
$\Phi^{(i)}_{\text{grid}}(\vec{x}) = \Phi^{(i)}_{\text{grid}}(\vec{x} + n \lambda_i), \quad \forall n \in \mathbb{Z}^2$.
However, the joint code across modules defines a unique representation of \( \vec{x} \) over a superlattice of volume:
$V_{\text{joint}} = \text{LCM}(\lambda_1, \dots, \lambda_M)$.
This multiscale module system enhances spatial resolution and disambiguation by reducing the conditional entropy \( H(\vec{x} \mid \{ \Phi^{(i)}_{\text{grid}} \}) \) via cross-scale phase comparisons and enabling robust localization and generalization through modular redundancy.
\end{corollary}
\textbf{Remark:} Under CCUP, the multiscale-module grid code \cite{stemmler2015connecting} provides a low-entropy, content-rich scaffold for reducing uncertainty in place-cell contexts, enabling scalable spatial inference for navigation \cite{bush2015using}. Additionally, multiscale-module grid code
supports error correction by resolving context aliasing through high-dimensional phase separation \cite{sreenivasan2011grid}.

\paragraph{Directional Symmetry Breaking by Place Cells.}
In the early stages of evolution, organisms operated in environments that were, from a sensorimotor perspective, spatially isotropic (no direction held inherent informational advantage over another). This symmetry poses a severe problem for adaptive behavior: in the absence of directional preference, every action is equally uncertain, and every percept equally ambiguous \cite{corballis2009evolution}. Without a mechanism to differentiate front from back, organisms cannot efficiently convert sensory input into meaningful, goal-directed behavior. The invention of \emph{steering} behavior constituted a fundamental \textbf{breaking in directional symmetry} \cite{bennett2023brief}. Organisms evolved bilateral bodies with sensory organs facing forward and propulsion mechanisms oriented backward. This architectural asymmetry allowed the organism to privilege information and action in the forward direction, reducing spatial uncertainty through active sampling and prediction \cite{friston2017active}. 

%\paragraph{Goal-Directedness as Spatial Inference.}
Steering does more than enable movement; it initiates \emph{spatial inference}. Once front and back are differentiated, the agent can begin to align its internal state with external gradients. The ability to steer toward or away from stimuli marks the origin of \textbf{goal-directed cognition} \cite{spreng2010default}: 1) goals emerge as \emph{preferred spatial locations} encoded in latent content variables \( \Phi \); 2) contexts \( \Psi \) are parsed as signals that inform steering decisions; 3) perception-action cycle enables the organism to iteratively reduce the conditional uncertainty between current state and goal state: \( H(\Phi \mid \Psi) \rightarrow 0 \) \cite{fuster2004upper}.
This transformation of space into a navigable, inferential medium reflects the first layer of cognitive abstraction: the world becomes not just a stimulus field, but a map with structure, gradients, and directionality \cite{whittington2022build}.
%\paragraph{Neural Implementation: From Directional Bias to Cognitive Maps.}
The emergence of steering behavior laid the foundation for spatial inference, but the evolution of two specialized neural representations greatly amplified its biological expression: 1) \textbf{head-direction (HD) cells} encode the animal's current heading within an allocentric reference frame \cite{taube1990head}. These cells provide essential \emph{context} variables \( \Psi \), establishing the directional prior that disambiguates spatial observations; 2) \textbf{place cells} in the hippocampus represent specific environmental locations and serve as low-entropy \emph{content} variables \( \Phi \) \cite{moser2008place}. 

\begin{proposition}[Unidirectional Place Cells for Directional Symmetry Breaking]
Let \( \Phi \) denote the content variable encoding spatial position, and \( \Psi \) the context variable encoding HD. In a symmetric environment, positions are invariant under rotational transformations:
$\forall \theta_1, \theta_2, \quad \Phi(x, y \mid \theta_1) = \Phi(x, y \mid \theta_2)$.
The emergence of unidirectional place cells breaks this directional symmetry by binding \( \Phi \) to HD-conditioned context \( \Psi \) s.t.
$\Phi(x, y \mid \theta_1) \neq \Phi(x, y \mid \theta_2)$.
%This symmetry breaking serves to disambiguate aliased contexts \( \Psi \) early in learning when sensory input is ambiguous.
\end{proposition}

\textbf{Remark:} Directional symmetry breaking increases representational specificity by encoding spatial positions as directionally modulated content and creating distinct place fields for HD-defined subspaces.
Initially, many place cells are \emph{unidirectional} (activate only when the animal passes through a location while facing a specific heading); they become omnidirectional through experience-driven generalization, which reflects the collapse of contextual aliasing in spatial memory \cite{buzsaki2013memory}.
%Under the CCUP framework, the interaction between HD cells and place cells illustrates a prototypical inference cycle: 1) Forward: HD cells (\( \Psi \)) disambiguate context to select the correct place-cell representation (\( \Phi \)), minimizing \( H(\Phi \mid \Psi) \); 2) Inverse: active place cells can be used to reconstruct HD states, minimizing \( H(\Psi \mid \Phi) \). 
Having initially broken directional symmetry to achieve context-specific spatial encoding, the system can now begin to generalize across directional contexts; this sets the stage for the emergence of omnidirectional place cells.

\paragraph{Generalization and the Emergence of Omnidirectional Place Cells.}

A key insight from CCUP is that the system must manage \textit{contextual aliasing} when different contexts \( \Psi \) produce similar content \( \Phi \). This is evident in how unidirectional place cells become \textbf{omnidirectional place cells} through repeated traversal (pp. 330, \cite{buzsaki2006rhythms}).
Initially, place cells encode spatial locations only under specific head directions. This is optimal for high-resolution disambiguation early in learning, but it is inefficient and brittle for generalization. As the animal explores an environment from multiple directions, its hippocampal system begins to consolidate aliased directional contexts into a single, unified place representation \cite{gyorgy2019brain}. That is:
$\Phi = f(\Psi_1) = f(\Psi_2) = \cdots = f(\Psi_n)
\quad \Rightarrow \quad H(\Phi \mid \Psi) \downarrow \text{ (generalization)}$.
%Through Hebbian learning and interference collapse, the system detects that multiple head-direction contexts \( \Psi_i \) consistently map to the same latent content \( \Phi \)—i.e., the same location. 
This generalization effectively collapses directional redundancy and reduces the dimensionality of the context space - i.e., $\delta(\Phi)=\int_{\Psi} \delta([\Phi,\Psi]) d\Psi$, where $\delta([\Phi,\Psi])$ and $\delta(\Phi)$ denote delta-seeded unidirectional and omnidirectional place cells, respectively. 
%In information-theoretic terms, omnidirectional place cells reflect a latent abstraction: multiple contextual inputs are merged into a single, low-entropy representation of place.

\paragraph{Summary.}
Spatial symmetry breaking begins with the formation of a general structure before specific encoding: grid cells provide a stable spatial scaffold that precedes and enables the formation of place cells. This reflects the broader principle that structure precedes specificity. In memory, the same spatial scaffold supports the consolidation of episodic memory into semantic memory, preserving the phylogenetic continuity between navigation and cognition \cite{buzsaki2013memory}. What begins as a physical asymmetry becomes a cognitive architecture for selective inference in space, bootstrapping more complex representational and planning capacities.

\begin{tcolorbox}[colback=gray!10!white, colframe=blue, title={\textbf{Spatial Cognition: Grid Cells Precede Place Cells}}]

%\textbf{Biological Insight:} \\
%Grid cells in the medial entorhinal cortex (MEC) exhibit periodic, hexagonally tiled spatial firing fields that form a universal spatial metric. Place cells in the hippocampus respond to specific environmental locations. Developmentally and evolutionarily, grid cells emerge earlier and provide the scaffolding upon which place-specific representations are constructed.

%\vspace{0.5em}

\textbf{Computational Principle:}
Grid cells encode a generalizable \textit{coordinate system}, a low-entropy structure across contexts.
Place cells bind \textit{specific experience} (high-resolution location) to this structure via contextual aliasing.
Therefore: structure precedes specificity \( H(\Phi_{\text{grid}}) < H(\Phi_{\text{place}} \mid \Psi) \), enabling stable inference across environments.

%\vspace{0.5em}

\textbf{Implication:} 
Grid cells reduce spatial uncertainty \( H(\Psi \mid \Phi_{\text{grid}}) \) by imposing regular structure, making contextual disambiguation possible. Place cells then refine this structure by binding specific content to localized context. Without the translational invariance of grid cells, place cells would suffer from aliasing and lack generalization across environments.

%\vspace{0.5em}

\textbf{Developmental Evidence:}
Grid cells appear earlier than place cells during postnatal development in rodents.
Inactivation of MEC disrupts place fields; inactivation of the hippocampus does not abolish grid fields.

%\textbf{Conclusion:} \\
%Grid cells implement the symmetry-breaking that creates representational geometry. Place cells are context-specific refinements built upon this scaffold, consistent with the CCUP strategy of “structure before specificity.”
\end{tcolorbox}

\section{Temporal Symmetry Breaking: Reinforcing and the Arrow of Inference}
\label{sec:4}

\paragraph{From Spatial Symmetry to Temporal Symmetry.}
While spatial symmetry breaking enables organisms to anchor themselves in the world, privileging front over back, and position over direction, it does not, by itself, support learning from experience. Once a stable spatial scaffold is in place, nature breaks the symmetry of time by introducing a directional bias: {\em the future becomes more behaviorally relevant than the past} \cite{schultz1997neural}. This temporal asymmetry is not hardwired but emerges through the evolutionary advantage of predicting outcomes based on past experiences and adapting behavior accordingly. Reinforcement learning, by assigning reward values to actions through temporally structured feedback, implements this broken symmetry, prioritizing forward associations over backward ones (pp. 292, \cite{buzsaki2006rhythms}). In this way, temporal symmetry breaking builds on the spatial scaffold, allowing agents to know both where they are and what to do next.

%Let \( \Phi_{\text{space}} \) denote low-entropy content variables encoding spatial structure (e.g., grid/place cell representations), and let \( \Phi_{\text{time}} \) denote content variables encoding temporally predictive structures (e.g., reward functions \cite{schultz1997neural}). Let \( \Psi \) denote high-entropy context variables corresponding to sensorimotor interactions and environmental transitions. Under CCUP, the emergence of temporal asymmetry in cognitive systems (i.e., the privileging of future-directed prediction over backward inference) requires a prior reduction of uncertainty in spatial representations. 
%\paragraph{From Temporal Symmetry to Predictive Bias.}
Just as early organisms encountered spatial ambiguity, they also faced temporal symmetry: without internal bias, the past and future are equally uncertain and indistinguishable in terms of their relevance to action. This symmetry is computationally intractable since organisms cannot afford to treat the entire temporal landscape as equally important \cite{minsky1961steps}. Dopamine-based \textbf{reinforcement learning} (RL) introduced a predictive bias into cognition, breaking this temporal symmetry by prioritizing \emph{future outcomes} based on \emph{past experience} \cite{schultz1997neural} (refer to the text box on NeuroAI). It established what we call the \emph{arrow of inference} \cite{li2025Arrow}: a structured asymmetry in time that enables prediction and control.
RL introduces a scalar signal (reward function), whose maximization requires learning reward functions oriented toward future states. Under CCUP, the past serves as a source of \emph{context} \( \Psi \), while the future is represented by \emph{goal-oriented content} \( \Phi \). The system learns a mapping from current observations and histories to expected future outcomes: i.e., predicting $E[\Phi|\Psi]$ via Rao-Blackwellization strategy \cite{blackwell1947conditional} instead of actual reward \cite{schultz1997neural}. Such predictive bias leads to the following result.

\begin{lemma}[RL as a Consequence of Spatial Anchoring]
Let \( V(\Phi_{\text{space}}) \) denote a value function over spatial content variables, representing expected future rewards conditioned on current position. Based on the bootstrapping order in Theorem \ref{thm:structure_precedes_specificity}, RL becomes tractable only when spatial representations are sufficiently structured:
given a structured spatial scaffold \( \Phi_{\text{space}} \), the agent can define temporally directed transitions:
    $\Phi_{\text{space}}(t) \rightarrow \Phi_{\text{space}}(t+1)$
    enabling the estimation of expected future value:
   $ V(\Phi_{\text{space}}) = \mathbb{E} \left[ \sum_{k=0}^{\infty} \gamma^k r_{t+k} \mid \Phi_{\text{space}}(t) \right]$, where $r_{t+k}$ denotes the future reward at $t+k$.
\end{lemma}

\textbf{Remark:} Without spatial symmetry breaking, state definitions are aliased, and the value function becomes ill-posed:
$\text{If } H(\Phi_{\text{space}}) \approx H(\Psi), \text{ then } V(\Phi_{\text{space}}) \text{ cannot converge reliably}$. This lemma links the temporal arrow of learning in RL to the spatial substrate over which it unfolds, consistent with both biological evolution and asymmetric inference. RL relies on the prior establishment of a low-entropy spatial state space via pattern recognition (PR) \cite{wehner1987matched}. Temporal credit assignment \cite{minsky1961steps} and prediction are meaningful only when transitions occur over a structured substrate of spatial representations, as shown by the following theorem (its proof can be found in Appendix C).

\begin{theorem}[Pattern Recognition as Spatial Symmetry Breaking]
Let \( \Psi \) denote high-entropy sensory input and \( \Phi_{\text{space}} \subset \Phi \) be a low-entropy latent representation extracted via pattern recognition. The initial observation space is spatially symmetric: all input configurations are equally probable.
Pattern recognition identifies structured content, breaking spatial symmetry and compressing observations:    $H(\Phi_{\text{space}}) \ll H(\Psi)$.
Under CCUP, this corresponds to contextual disambiguation that minimizes conditional entropy:  
    $H(\Phi_{\text{space}} \mid \Psi) \to \min$.
%Therefore, pattern recognition acts as a mechanism of spatial symmetry breaking, transforming ambiguous, unstructured input into structured content suitable for inference and action.
\end{theorem}

\paragraph{From Spatial and Temporal Symmetry Breaking to the Emergence of a World Model.}
The formation of a world model arises from the coordinated interaction between PR and RL, each contributing a complementary form of symmetry breaking under CCUP. PR reduces the entropy of high-dimensional observations by identifying spatially invariant features and extracting structured latent representations \cite{wehner1987matched}. This process breaks spatial symmetry by selecting a subset of meaningful configurations from the raw sensory manifold, thereby transforming context into interpretable content \cite{serre2007robust}. RL then builds upon this spatial scaffold by interacting with the environment to predict temporally extended outcomes. Through trial and error, RL breaks temporal symmetry by assigning value to specific transitions, prioritizing future/goal-directed inferences over past or reversible ones. As PR stabilizes content and RL injects contextual dynamics, their integration enables the emergence of a predictive, generative world model \cite{ha2018world}, which we formalize below. %Within this model, spatial representations provide anchors for temporal simulation, and the inference loop between perception and prediction minimizes conditional entropy in both directions. The result is a cycle-consistent system that binds relevant features into latent content while disentangling context through active interaction with the environment.

\begin{definition}[World Model]
A \emph{world model} is a structured generative system that learns a latent representation \( \Phi \) of observations \( \Psi \) and predicts future states and rewards under action sequences. Formally, a \textbf{world model} is a tuple
$\mathcal{M} = \left( \phi, \mathcal{T}, \mathcal{R} \right)$,
where: \( \phi: \Psi \rightarrow \Phi \) is an encoding function that maps sensory observations \( \Psi \) to a latent variable \( \Phi \), extracting structured content from ambiguous context; \( \mathcal{T}: \Phi \times \mathcal{A} \rightarrow \Phi \) is a transition model that predicts future latent states: \( \Phi_{t+1} \sim \mathcal{T}(\Phi_t, a_t) \); \( \mathcal{R}: \Phi \times \mathcal{A} \rightarrow \mathbb{R} \) is a reward model that estimates the expected value $V$ of executing action \( a_t \) in latent state \( \Phi_t \).
\end{definition}

\paragraph{Acquisition of a World Model via PR and RL.} The world model enables the agent to simulate and evaluate counterfactual action sequences and serves as the basis for model-based planning and active inference \cite{friston2017active}. Under CCUP, encoding function \( \phi \) resolves the context-content asymmetry by minimizing conditional entropies, 
thereby establishing a cycle-consistent representation that supports temporally directed learning. The construction of a world model emerges through the cooperative interaction between PR and RL \cite{ha2018world}. The agent begins with raw sensory observations that are high in entropy due to ambiguity and redundancy. PR acts as a contextual disambiguator by extracting stable latent structures, such as objects, landmarks, or perceptual features, that compress the high-dimensional input into a low-entropy content space \cite{dicarlo2012does}. This process breaks spatial symmetry and enables the identification of meaningful anchors in the latent state space. RL then bootstraps temporal inference on top of this spatial structure, using goal-directed interaction \cite{spreng2010default} to learn transition dynamics and reward contingencies. Temporal symmetry is broken by the asymmetry of reward-based evaluation, which biases the system toward predicting and optimizing future outcomes \cite{bastos2012canonical}. Over time, the bidirectional interaction between perception and prediction forms a cycle: PR refines latent encodings from sensory input, while RL uses these encodings to generate expectations that guide behavior. This cycle minimizes conditional entropies in both directions, \( H(\Phi \mid \Psi) \) and \( H(\Psi \mid \Phi) \), thereby generating a cycle-consistent world model as follows. %Within this framework, content becomes structured through selective inference, and context is disentangled through dynamic consistency, yielding a predictive, generative representation of the environment.

\begin{proposition}[Temporal World Modeling]
Let \( \Psi \) denote raw high-dimensional sensory observations, and let \( \Phi \) denote latent representations extracted via PR. Let \( \Phi_{\text{space}} \subset \Phi \) denote spatially organized representations, and \( \Phi_{\text{time}} \subset \Phi \) denote temporally inferred representations learned through RL. Then the world model emerges as a cycle-consistent mapping between content \( \Phi \) and context \( \Psi \), in which spatial structure serves as the scaffold for temporal prediction: 1) spatial symmetry is first broken through PR:    $H(\Phi_{\text{space}}) \ll H(\Psi)$;
    2) temporal inference is bootstrapped by spatial anchoring:
    $H(\Phi_{\text{time}} | \Phi_{\text{space}}) \ll H(\Phi_{\text{time}} | \Psi)$;
    3) RL-induced inference cycle reduces conditional entropy in both directions:
    $H(\Phi | \Psi) + H(\Psi | \Phi) \to \min$.   
\label{thm:world_model}
\end{proposition}

\textbf{Remark:} This proposition emphasizes that temporal prioritization, learning from future-directed prediction, is essential not only for reward learning but for the emergence of structured world models that support intelligent behavior. Temporal symmetry breaking via RL is a necessary precondition for world model acquisition \cite{wu2025rlvr}. The directed flow of predictive error enables the construction of a low-entropy internal transition model, which in turn serves as the foundation for goal-directed simulation in the latent space, as we show next.

\begin{corollary}[World Model Enables Simulation via Inference Cycles]
Let \( T: (\Phi_t, a_t) \mapsto \Phi_{t+1} \) denote a learned world model acquired through Proposition \ref{thm:world_model}. This internal model can be reused for goal-directed simulation by iterating inverted inference cycles in latent space as follows.
1) \textbf{Content-to-Context Prediction:} The world model generates forward predictions:
    $\Phi_t, a_t \rightarrow \Phi_{t+1}$
    simulating future states conditioned on current content and action;
2) \textbf{Context-to-Content Inference:} Inverted/active inference selects actions that align predicted content with goal-conditioned context:
    $\Psi_{\text{goal}} \rightarrow \Phi_t,~\text{choose } a_t \text{ such that } T(\Phi_t, a_t) \approx \Psi_{\text{goal}}$;
3) \textbf{Closed Inference Loop:} The system alternates between:
    $\Phi_t \rightarrow \Psi_t \rightarrow \Phi_{t+1} \rightarrow \Psi_{t+1} \rightarrow \cdots$
    refining both action selection and state prediction through iterative minimization of
    $H(\Psi_{t+k} | \Phi_{t+k}) + H(\Phi_{t+k} | \Psi_{t+k})$.
\end{corollary}

\textbf{Remark:} This corollary captures the shift from reactive learning to generative simulation (a.k.a. ``inverted inference'' or ``active inference'' \cite{friston2017active}), where the same structures learned through temporal prediction are $repurposed$ to simulate futures, compare with goals, and select optimal actions. Consistent with the structure-before-specificity principle, the world model closes the inference loop by predictive coding \cite{keller2018predictive}, allowing the agent to perceive and act under uncertainty. This recursive use of internal dynamics fulfills the CCUP requirement of reducing uncertainty across context-content transitions through goal-directed simulation \cite{pezzulo2014internally}.
Building upon the cycle-consistent structure of the world model, the agent’s ability to simulate future trajectories naturally gives rise to a valuation mechanism in which temporal preferences emerge not solely from external reward signals but from the internal drive to minimize uncertainty over extended horizons, thereby linking temporal discounting to an entropy-weighted reward mechanism \cite{li2025multi}.

\paragraph{Discounting and the Compression of Time.}
A defining feature of RL is \emph{temporal discounting} \cite{sutton1998reinforcement}, implemented via a factor \( \gamma \in (0,1) \), which devalues rewards in the distant future. This prioritizes near-future outcomes over distant ones, effectively compressing the temporal horizon into a low-entropy content space:
$V(s) = \mathbb{E}\left[ \sum_{t=0}^{\infty} \gamma^t r_{t} \,\middle|\, s_0 = s \right]$.
This compression reduces the complexity of learning and reflects an adaptive trade-off: short-term predictability is weighted more heavily than long-term speculation. From the CCUP perspective, temporal discounting minimizes \( H(\Phi \mid \Psi) \) by reducing variance in future content predictions, allowing learning systems to stabilize inference within the bounds of uncertainty. We formalize the above intuition below.

\begin{theorem}[Temporal Discounting as Entropy-Weighted Reward]
Let \( \Phi_t \) denote low-entropy content variables encoding latent state at time \( t \), and let \( r_{t+k} \) denote reward received at future time \( t+k \). In RL, the expected return is typically computed via a discounted sum:
$V(\Phi_t) = \mathbb{E} \left[ \sum_{k=0}^{\infty} \gamma^k r_{t+k} \right], \quad \gamma \in [0,1)$
Under CCUP, temporal discounting \( \gamma^k \) reflects an adaptive weighting of future rewards based on increasing conditional entropy over time. Formally, if the conditional entropy of future content increases with temporal distance: $\frac{\partial H(\Phi_{t+k} \mid \Phi_t)}{\partial k} > 0$ and uncertainty-modulated reward is defined as:
    $V^\ast(\Phi_t) = \sum_{k=0}^{\infty} w_k \cdot \mathbb{E}[r_{t+k}]
    \quad \text{with} \quad w_k \propto \frac{1}{H(\Phi_{t+k} \mid \Phi_t)}$, then the standard discounting scheme \( \gamma^k \) is an approximation to entropy-weighted reward under a monotonic uncertainty prior:
$\gamma^k \approx \exp\left(-\lambda H(\Phi_{t+k} \mid \Phi_t)\right), \lambda > 0$.
\end{theorem}

\textbf{Remark:} This theorem connects the commonly used geometric discounting in RL to a deeper principle of entropy-weighted planning \cite{zhao2020weighted}: it shows that predictive uncertainty justifies devaluation of the distant future, rather than assuming time preference arbitrarily. Temporal discounting is not arbitrary: it approximates a rational valuation scheme that prioritizes rewards based on their inferential certainty. As uncertainty about future states grows, their impact on present value should decrease accordingly \cite{watkins1992q}. This formalizes discounting as a direct computational consequence of temporal symmetry breaking under CCUP. Having established temporal discounting as a principled response to growing uncertainty over time, we now turn to a complementary dynamic: how agents regulate their behavior between exploiting known rewards and exploring uncertain contexts \cite{sutton1998reinforcement}, based on the direction and magnitude of entropy gradients between content and context?

\begin{corollary}[Exploration–Exploitation Tradeoff as Entropy Gradient Alignment]
Let \( \Phi \) denote latent content representations and \( \Psi \) denote high-entropy context variables in the environment. An intelligent agent manages uncertainty by alternating between:
1) \textbf{Exploitation:} Reducing \( H(\Psi \mid \Phi) \) by selecting actions \( a \) that align current internal representations with familiar external contexts to maximize expected reward; and   
2) \textbf{Exploration:} Reducing \( H(\Phi \mid \Psi) \) by selecting actions that expose the agent to ambiguous contexts to refine its content representations.
Let the entropy gradients be defined as:
$\nabla_{\text{exploit}} := \frac{\partial H(\Psi \mid \Phi)}{\partial a},
\nabla_{\text{explore}} := \frac{\partial H(\Phi \mid \Psi)}{\partial a}$.
The optimal exploration–exploitation balance corresponds to a local policy \( \pi^* \) that dynamically minimizes total expected uncertainty:
$\pi^*(a) = \arg\min_a \left[ H(\Psi \mid \Phi) + H(\Phi \mid \Psi) \right]$.
\end{corollary}

\textbf{Remark:} This corollary reframes exploration and exploitation as complementary inference strategies, dynamically governed by the direction of uncertainty gradients between context and content. The exploration–exploitation tradeoff \cite{audibert2009exploration} is a dynamic alignment strategy guided by CCUP. The agent shifts behavior based on which entropy gradient dominates, exploring when internal representations are underspecified and exploiting when context is confidently mapped to content. While the exploration–exploitation tradeoff governs how agents manage uncertainty relative to external rewards, curiosity corresponds to a drive to reduce internal uncertainty where content–context alignment is weak. Accordingly, {\em curiosity-driven learning} \cite{pathak2017curiosity} emerges when the agent seeks to reduce uncertainty for its own sake, actively selecting experiences that maximize learning progress through intrinsic entropy minimization. This leads to the following corollary.

\begin{corollary}[Curiosity-Driven Learning as Intrinsic Entropy Minimization]
Let \( \Phi \) denote internal content representations and \( \Psi \) denote external sensory context. 
Let the agent's curiosity signal \( \mathcal{C}(\Psi) \) be defined by the rate of change in prediction error or uncertainty:
$\mathcal{C}(\Psi) := -\frac{d}{dt} H(\Phi \mid \Psi)$.
An agent exhibits curiosity-driven behavior when it selects actions that maximize expected reduction in representational uncertainty:
$a^* = \arg\max_a \mathbb{E}\left[ \mathcal{C}(\Psi_{t+1}) \right] = \arg\max_a \left[ H(\Phi \mid \Psi_t) - H(\Phi \mid \Psi_{t+1}) \right]$.
\end{corollary}

\textbf{Remark:} In curiosity-driven learning, agents are intrinsically motivated to seek experiences that maximize learning progress, not just extrinsic reward. Curiosity, acting as a self-supervised objective for discovering latent structure in the environment, emerges as an internal implementation of CCUP, driving agents to explore regions of high uncertainty that can be reduced through learning \cite{pathak2017curiosity}. This intrinsic entropy minimization promotes efficient representation building even in the absence of external rewards, which results in the following behavior: the agent seeks novel or ambiguous \( \Psi \) where \( H(\Phi \mid \Psi) \) is initially high and avoids both fully predictable \( \Psi \) (no learning gain) and fully random \( \Psi \) (no structure to learn).
In sum, curiosity, an active inference strategy that selects actions to maximize the rate of entropy reduction,  arises naturally from the mismatch between context and content and works to close that gap over time.

\paragraph{Summary.}
Temporal symmetry breaking prioritizes future prediction over past observation, enabling reinforcement learning to assign value, guide behavior, and build structured world models. This forward-directed inference reflects the principle of structure-before-specificity in time, where agents first learn predictive structure before optimizing specific actions. By weighting temporally distal outcomes and learning from uncertainty reduction, cognition transforms time into a scaffold for abstraction. Under CCUP, temporal asymmetry aligns anticipated context with selected content, minimizing uncertainty through directed inference.

\begin{tcolorbox}[colback=gray!10!white, colframe=blue, title={\textbf{NeuroAI: Dopamine as the Biological Substrate of Reinforcement Learning}}]

%\textbf{Biological Insight:} \\
%Dopamine neurons, especially in the \textit{ventral tegmental area (VTA)} and \textit{substantia nigra}, encode a \textit{reward prediction error (RPE)}---the difference between expected and actual outcomes. This neural signal mirrors the temporal difference (TD) error used in reinforcement learning (RL).

\vspace{0.5em}

\textbf{Computational Mapping:}

\begin{center}
\begin{tabular}{@{}l l@{}}
\toprule
\textbf{RL Concept} & \textbf{Neural Substrate} \\
\midrule
State \( s_t \) & Cortical + hippocampal input \\
Action \( a_t \) & Motor intention / basal ganglia output \\
Reward \( r_t \) & Sensory or affective signal (e.g., food, safety) \\
Value Function \( V(s) \) & Ventral striatum, orbitofrontal cortex \\
TD Error \( \delta_t = r_t + \gamma V(s_{t+1}) - V(s_t) \) & Phasic dopamine response \\
\bottomrule
\end{tabular}
\end{center}

\vspace{0.5em}

\textbf{Key Roles of Dopamine:}
\begin{itemize}
    \item Signals \textit{surprise} and \textit{novelty}, driving adaptive learning.
    \item Encodes the \textit{direction and magnitude} of behavioral updates.
    \item Modulates \textit{synaptic plasticity}, especially in corticostriatal circuits.
\end{itemize}

\vspace{0.5em}

\textbf{Implication under CCUP:} \\
Dopamine functions as a neuromodulatory messenger that minimizes temporal uncertainty \( H(\Phi_{\text{time}} \mid \Psi) \). It links past context to future value, enabling \textit{directional asymmetry} in time by aligning expectations with sensory input.

\end{tcolorbox}

\section{Separation of Self from World and Internal Asymmetry}
\label{sec:5}

\paragraph{Self-World Separation via Structure-Before-Specificity.}
The separation between self and world arises as a structured decomposition of sensory input, guided by the principle of structure-before-specificity (Theorem \ref{thm:structure_precedes_specificity}). Initially, the agent receives high-entropy observations that conflate external (world-driven) and internal (self-driven) causes. PR first identifies stable, low-entropy regularities in the environment, giving rise to a latent world model \( \Phi_{\text{world}} \) that captures generalizable structure independent of specific actions. This inference breaks spatial and temporal symmetry, forming the foundational scaffold for further prediction \cite{ha2018world}. The notion of self, \( \Phi_{\text{self}} \), then emerges as the residual source of variance, the internal cause of deviations from the expected dynamics of the external world. In this sense, the self is inferred only after the world has been stabilized, as the system explains away prediction errors not accounted for by the world model \cite{wu2025rlvr}. This hierarchy of inference minimizes the joint conditional entropy over sensory input, such that \( H(\Phi_{\text{world}}) \ll H(\Phi_{\text{self}}) \) and \( H(\Psi \mid \Phi_{\text{world}}, \Phi_{\text{self}}) \to \min \). 
The self-world separation reflects a principled way of distilling latent causes, where structural inference over the environment precedes and conditions the emergence of self-specific dynamics \cite{xie2024making}.

Once the separation between self and world has been established through structured inference, a deeper form of internal asymmetry emerges: {\em the ability to simulate future outcomes based on internally specified goals}, which marks a critical transition from passive perception to active inference \cite{friston2017active}. The agent, having distinguished its latent states \( \Phi_{\text{self}} \) from external dynamics \( \Phi_{\text{world}} \), can now reconfigure its internal dynamics to prioritize certain outcomes over others. This process introduces an asymmetry not imposed by the environment but seeded internally, giving rise to directed simulations that bias action selection and planning \cite{spreng2010default}. In this setting, goals act as low-entropy anchors that selectively condition inference and compress future trajectories into meaningful paths. Temporal symmetry is thus broken not merely by environmental contingencies but by the agent's asymmetric valuation of possible futures. The combination of self and world modeling enables model-based RL to reframe the search problem as goal-directed inference, achieving efficient planning through structure-aware, specificity-driven simulation \cite{moerland2023model}. %This \textbf{internal asymmetry} enables the system to convert structured world knowledge into purpose-driven behavior, establishing a generative loop in which imagined futures guide present actions, and actions refine future expectations. Goal simulation, therefore, extends the structure-before-specificity principle into the temporal and motivational domain, transforming inert self-world structure into anticipatory control.

%\paragraph{Goal Simulation as Internal Asymmetry.}
\paragraph{Model-based RL Solves the Search Problem.}
While spatial and temporal symmetry breaking introduce directional biases that stabilize representation and prioritize future-oriented inference, they do not by themselves resolve the core challenge of adaptive behavior: \emph{how to efficiently search through the vast space of possible future actions to achieve desired outcomes} \cite{minsky1961steps}. This has been referred to as the "search problem" by Marvin Minsky \cite{minsky1986society}, a computational bottleneck arising from exponential representational complexity. In a symmetric spacetime landscape, all trajectories are initially equiprobable, and without internal constraints, behavioral exploration becomes computationally intractable, especially in high-dimensional space (e.g., due to the curse of dimensionality \cite{bellman1966dynamic}).
The evolution of \textbf{goal simulation} addressed this challenge by introducing an internal asymmetry between the \emph{agent} and the \emph{environment}. Rather than relying solely on reactive responses to sensory input, the agent gains the ability to internally simulate potential actions and predict their consequences before committing to external execution. 
%This capability marks a critical cognitive innovation: the emergence of a representational distinction between a \textbf{self-model} (an internal generative system encoding latent goals, intentions, and hypothetical trajectories) and a \textbf{world-model}, which captures the statistical structure and contingencies of the external environment.
By temporally decoupling inference from action, goal simulation enables the agent to anticipate, evaluate, and optimize future behavior through counterfactual reasoning \cite{pearl2009causality}. This \textbf{internal asymmetry}, an epistemic separation between simulation and execution, supports the development of planning, imagination, and lays the foundation for higher-order cognitive functions \cite{block1995confusion}. Goal simulation, therefore, extends the structure-before-specificity principle into the imaginary and motivational domain, transforming inert self-world structure into anticipatory control.

More specifically, goal simulation breaks spatiotemporal symmetry by creating a privileged internal pathway from imagined content \( \Phi \) (goal states) to inferred context \( \Psi \) (conditions under which those goals are achievable). This allows the brain to $invert$ the typical inference flow (e.g., predictive coding \cite{keller2018predictive} and active inference \cite{friston2017active}):
$\underbrace{\Psi \rightarrow \Phi}_{\text{perception}}  \longleftrightarrow 
\underbrace{\Phi \rightarrow \Psi}_{\text{simulation}}$ for breaking the curse of dimensionality.
Traditional inference cycle consists of 1) \textbf{forward (perception)}: disambiguate context \( \Psi \) to infer current state and possible goals \( \Phi \); 2) \textbf{backward (simulation)}: generate desired future content \( \Phi \) and infer the context \( \Psi \) to realize it.
The inverted cycle reverses the ordering by pre-emptively generating predictions of what it expects to observe and only updates its internal beliefs when there is a mismatch. This inverted inference cycle \cite{keller2018predictive}, seeded by internally specified goals, establishes a directional bias in the agent’s predictive machinery, transforming symmetric latent dynamics into asymmetric control policies that prioritize futures consistent with goal-conditioned expectations. Formally, we have

\begin{theorem}[Inverted Inference Facilitates Goal-Directed Simulation]
Let \( \Phi_{\text{goal}} \) be a low-entropy goal state in the latent space, and let \( \Phi_{0:T} \) be a trajectory of latent states inferred by simulation. Then inverted inference enables efficient goal simulation by:
1) {reversing the direction of inference:}  
    $p(\Phi_{0:T} | \Phi_{\text{goal}}) \ll p(\Phi_{0:T} | \Phi_0)$;   
2) {minimizing conditional entropy of the simulated path:}
    $H(\Phi_{0:T} | \Phi_{\text{goal}}) \ll H(\Phi_{0:T} | \Phi_0)$;
3) {restricting inference to goal-relevant trajectories:}
    $\Phi_{0:T} \in \mathcal{M}_{\text{goal}}, \dim(\mathcal{M}_{\text{goal}}) \ll \dim(\Phi)^T$.
\end{theorem}

%\begin{theorem}[Self-World Modeling Enables Model-Based RL to Solve the Search Problem]
%Let \( \Phi_{\text{world}} \) and \( \Phi_{\text{self}} \) be learned latent representations of environmental dynamics and agent-specific variables, respectively. Let \( \Phi_{\text{goal}} \subset \Phi_{\text{self}} \) denote an internally specified target state. Under a model-based RL framework with goal-conditioned inference, the agent simulates trajectories in latent space using its generative model: $(\Phi_t, a_t) \xrightarrow{\mathcal{T}} \Phi_{t+1}, \quad \text{with } \Phi_0 = \phi(\Psi_0)$. The search problem over action sequences becomes a constrained inference problem: $\text{Infer } \{a_0, \ldots, a_T\} \text{ such that } \Phi_T \approx \Phi_{\text{goal}}$. The effective search space is restricted to a low-dimensional goal manifold: $\mathcal{O}(\text{search}) \ll |\mathcal{A}|^T, \text{due to structural guidance from } \Phi_{\text{world}} \text{ and } \Phi_{\text{self}}$.
%\end{theorem}

\textbf{Remark:} The proof of the above theorem is referred to Appendix E.
Inverted inference transforms the task of goal simulation from an intractable forward search problem into a structured inference process over a subset of the latent state space. Under the inverted inference view, the agent simulates not every possible future but only those that are consistent with a low-entropy target. The goal functions as a structural prior, much like a reward function in reinforcement learning, but with stronger conditioning on latent content. This formulation aligns with the backward recursion in dynamic programming~\cite{bellman1966dynamic} while extending it to inference in generative latent spaces.
Importantly, inverted inference does not negate the utility of value functions or learned models; rather, it complements them by providing a mechanism for top-down simulation that integrates structural knowledge and internal motivation. In doing so, inverted inference generalizes the principle behind the half-step trick in Q-learning, applying it not just at the level of immediate transitions, but across entire trajectories.

\paragraph{Goal-directed Simulation with Episodic Memory.}

The agent must now engage in an exploration-exploitation cycle, wherein episodic memory plays a critical role in guiding asymmetric (goal-directed) control by balancing the discovery of new structure with the selective reuse of past experience.
The core insight behind goal-directed control is that intelligent agents do not merely react to external stimuli but actively shape their inference and behavior through internally simulated goals. This introduces a fundamental asymmetry in the inference process: rather than passively interpreting observations, {\em the agent uses its self-model to generate content-hypothetical futures and latent goals}, which then constrain the interpretation of context \cite{friston2017active}. This directionality transforms the search problem into a goal-conditioned inference process, in which exploration is structured by prior intentions and exploitation is grounded in predictive evaluation \cite{sutton1998reinforcement}. Under CCUP, this form of control minimizes uncertainty not by brute-force sampling but by selectively aligning internally simulated content with externally constrained context \cite{li2025Arrow}. Asymmetric control thus redefines the exploration–exploitation tradeoff \cite{sutton1998reinforcement} as an inference cycle driven by internal goals and refined by contextual feedback, enabling tractable decision-making in high-dimensional, uncertain environments. To formalize this line of reasoning, we start with a construction of an inference cycle for asymmetric control.

\begin{definition}[Exploration–Exploitation Cycle for Asymmetric Control]
Let \( \Phi \) represent internal content (e.g., goal simulations) and \( \Psi \) represent external context (e.g., sensory input). The agent alternates between 1) \textbf{Exploration:} Disambiguating \( \Psi \) to inform or adjust \( \Phi \), minimizing \( H(\Phi | \Psi) \);
  and 2) \textbf{Exploitation:} Projecting \( \Phi \) into \( \Psi \), predicting and acting to confirm \( \Phi \), minimizing \( H(\Psi | \Phi) \).
This cycle breaks the symmetry of undirected exploration and transforms inference into an active, goal-conditioned process.
\end{definition}

Following Friston's free energy formulation \cite{friston2010free}, we have the following lemma for the newly defined exploration–exploitation cycle.

\begin{lemma}
    [Variational Inference Control for Exploration–Exploitation Tradeoff]
Let \( \Psi \) be external context (observations), \( \Phi \) be internal content (simulated goals, latent states), and \( q(\Phi) \) be the agent’s posterior belief over latent states. Let \( p(\Psi, \Phi) = p(\Psi | \Phi)p(\Phi) \) define the generative model. The agent minimizes the variational free energy:
$\mathcal{F}[q] = \underbrace{\mathbb{E}_{q(\Phi)}[-\log p(\Psi | \Phi)]}_{\text{Context disambiguation}} + \underbrace{\text{KL}(q(\Phi) \| p(\Phi))}_{\text{Content complexity}}$.
The exploration–exploitation tradeoff between prediction error and exploitation cost is thus governed by the balance
between exploration: $\min H(\Phi | \Psi)$ \text{(reduce ambiguity by updating content)} and exploitation: $\min \text{KL}(q(\Phi) \| p(\Phi))$ \text{(stay close to generative priors)}.
\end{lemma}

\textbf{Remark:} The exploration–exploitation cycle converges when both prediction error and content deviation are minimized, aligning internal goals with external contingencies. From the perspective of dynamic programming \cite{bellman1966dynamic}, we note that this lemma implies that the exploration–exploitation tradeoff is not a toggle but
a gradient flow in the variational free energy landscape, governed by entropy and complexity. At the heart of the exploration-exploitation cycle lies episodic memory, which serves as a dynamic repository of specific experiences that enables the agent to balance novelty-driven exploration with goal-directed exploitation through selective retrieval and context-sensitive reuse of past interactions.

\paragraph{Dual Role of Episodic Memory.} Episodic memory \cite{tulving2002episodic} provides the mechanism by which past-specific experiences are stored and reused to inform future action, allowing the agent to bias exploitation while still supporting efficient exploration asymmetrically. On the one hand, episodic memory transforms the exploration-exploitation tradeoff into a goal-conditioned, asymmetrically biased control cycle, enabling selective reuse of specific past experience while dynamically identifying regions of uncertainty for exploration. On the other hand, episodic memory enables the agent to bootstrap goal-driven inference with concrete, uncertainty-reducing priors derived from past experience. We summarize the dual role of episodic memory into the following two propositions.

\begin{proposition}[Episodic Memory Supports Goal-Driven Inference]
Let \( \mathcal{M}_{\text{epi}} \) be the episodic memory store containing prior experience traces \( \{\tau_i\} \), each mapping context \( \Psi_i \) to outcome-relevant content \( \Phi_i \). Let the agent engage in goal-driven inference to simulate trajectories \( \Phi_g \) toward a latent goal \( g \). Then 1) Retrieval from \( \mathcal{M}_{\text{epi}} \) reduces simulation entropy: $H(\Phi_g | g, \mathcal{M}_{\text{epi}}) \ll H(\Phi_g | g)$; 2) Retrieved episodes constrain context–content alignment by populating the inference cycle with grounded trajectories.
\end{proposition}

\begin{proposition}[Episodic Memory Enables Asymmetric Control in the Exploration-Exploitation Cycle]
Let \( \mathcal{E} = \{ (\Psi_t, a_t, \Psi_{t+1}, r_t) \} \) denote a memory bank of episodic traces, and let \( \Phi_{\text{goal}} \) represent a currently active, internally seeded goal. Then we have 1) exploitation via episodic matching:
$\Phi_{\text{goal}} \xrightarrow{\text{nearest-neighbor match}} (\Psi_{t}, a_{t}, r_{t}) \in \mathcal{E}, \text{enabling rapid reuse of past actions.}$; 2) exploration via episodic novelty detection:
$\text{If } \nexists \, (\Psi_{t}, a_{t}) \approx \Phi_{\text{goal}}$, then explore: sample new transitions;
3) asymmetric control via goal-biased retrieval:
$\text{Memory retrieval is not uniform but biased toward episodes with } \max r_t \text{ consistent with } \Phi_{\text{goal}}.$
\end{proposition}

\textbf{Remark:} Episodic memory plays a dual role in adaptive intelligence: on one hand, it supports goal-driven inference by enabling the retrieval of content anchored to internal states and enables asymmetric control by biasing action selection toward reward-relevant, context-specific past experiences. On the other hand, episodic memory's integration into model-based inference systems enables the construction of temporally extended control hierarchies \cite{badre2008cognitive}. 

\paragraph{Model-based Hierarchical RL for Embodied Cognition.}
Episodic traces serve not only as memory units but as building blocks for simulating multi-step, goal-aligned trajectories. These traces can be composed, abstracted, and reorganized into higher-level plans that coordinate sensorimotor interaction across spatial and temporal scales \cite{fuster2004upper}. This compositional structure gives rise to model-based hierarchical control \cite{merel2019hierarchical}, wherein low-level motor primitives are guided by mid-level routines and high-level intentional frames, each constrained by distinct levels of abstraction within the world and self models \cite{bennett2023brief}. Importantly, this hierarchical organization is grounded in embodiment \cite{shapiro2019embodied}: the agent’s physical constraints and affordances shape both the content of episodic memory and the structure of inferred plans. Thus, episodic memory bridges reactive control and strategic planning, enabling embodied agents to act with purpose, adaptivity, and coherence in complex environments.

\begin{proposition}[Recursive Tractability in Model-Based Hierarchical RL]
Let the agent maintain \( L \) levels of abstraction, where each level \( l \in \{1, \dots, L\} \) contains:
1) a self-model \( \mathcal{G}_{\text{self}}^{(l)} \) generating goal-conditioned trajectories over abstract actions \( \mathcal{A}^{(l)} \); 2) a world-model \( \mathcal{G}_{\text{world}}^{(l)} \) predicting transition dynamics \( p^{(l)}(s' \mid s, a) \); and 3) a dynamic programming routine computing local reward functions \( V^{(l)}(s) \) and reward functions at level \( l \) are used to guide planning at level \( l+1 \).
If each level restricts planning to a goal-consistent subspace:
$\mathcal{T}^{(l)}_{\text{self}} = \left\{ \tau \in \mathcal{T}^{(l)} : \tau \sim \mathcal{G}_{\text{self}}^{(l)}(g) \text{ and } \mathbb{E}[V(\tau)] \geq \theta^{(l)} \right\}$, then the total complexity of planning reduces from:
$\mathcal{O}(|\mathcal{A}|^T) \quad \text{to} \quad \sum_{l=1}^L \mathcal{C}^{(l)} \cdot \text{depth}^{(l)}$,
where \( \mathcal{C}^{(l)} \) is the cost at level \( l \), and \( \text{depth}^{(l)} \ll T \).
\end{proposition}

\textbf{Remark:} Recursive self-world separation and dynamic programming across hierarchical abstraction transforms exponentially hard planning into a tractable, layered inference process. This is consistent with the evolution of hierarchical goals (i.e., goal $\rightarrow$ subgoal $\rightarrow$ subsubgoal) in mammalian brains \cite{bennett2023brief}. Formally, we have the following corollary.

\begin{corollary}[Hierarchical Planning for Model-based RL]
Let \( \{(\Psi^{(l)}, \Phi^{(l)})\}_{l=1}^L \) denote a hierarchy of context-content pairs across \( L \) levels of abstraction, with \( \Phi^{(l)} \) denoting content at level \( l \), and \( \Psi^{(l)} \) denoting its corresponding context.
Hierarchical planning proceeds by minimizing total conditional uncertainty across levels via recursive inference:
$\sum_{l=1}^{L} \left[ H\left(\Phi^{(l)} \mid \Psi^{(l)}\right) + H\left(\Psi^{(l)} \mid \Phi^{(l)}\right) \right] \rightarrow \min$
subject to cross-level consistency constraints:
$\Phi^{(l)} = f^{(l)}\left(\Phi^{(l+1)}\right), \Psi^{(l+1)} = g^{(l)}\left(\Psi^{(l)}\right)~ \text{for } l = 1, \dots, L-1$,
where \( f^{(l)} \) and \( g^{(l)} \) denote top-down abstraction and bottom-up contextualization mappings, respectively.
This structure enables the agent to disambiguate high-level goals \( \Phi^{(L)} \) into concrete plans \( \Phi^{(1)} \) via successive context reconstruction, generalize from low-level sensorimotor contexts \( \Psi^{(1)} \) to abstract contextual priors \( \Psi^{(L)} \),
and maintain dynamic consistency across all levels through a cascade of exploration-exploitation cycles.    
\end{corollary}

\textbf{Remark:} The optimality of goal-directed simulation is grounded in the principle of dynamic programming, which asserts that globally optimal solutions can be constructed from recursively optimal subproblems~\cite{bellman1966dynamic}. In classical formulations, dynamic programming proceeds by forward rollout or backward value iteration across a state–action tree. Goal-directed simulation preserves this recursive optimality but inverts the direction of computation: rather than expanding from the current state forward, it conditions on a low-entropy internal goal and infers backward the most probable trajectory that could have led to it. This inversion reframes Bellman's principle not as a forward search, but as a constrained inference process over latent paths \cite{watkins1992q}. The recursive structure of dynamic programming remains intact: each subgoal must still be optimally achievable given its successor, but the computation is seeded from the top of the goal hierarchy and propagated downward through generative models. %As such, GDS achieves the same optimal substructure guarantees while reducing computational complexity through goal anchoring and hierarchical abstraction.

% Add this where you want the sidebar to appear
\begin{tcolorbox}[colback=gray!5!white,colframe=blue!75!white,title= Structural Analogy: Bootstrapped Q-Learning vs. Inverted Inference]

\textbf{Core Idea:} 

Both Q-learning and inverted inference address the curse of dimensionality by avoiding full expansion of the action-state tree, but they do so through complementary mechanisms: 

1) \textbf{Q-learning} uses \emph{temporal bootstrapping}:
    $Q(s, a) \leftarrow r + \gamma \max_{a'} Q(s', a')$, which propagates value without computing full future returns, reducing computational cost; 
    
2) \textbf{Inverted inference} uses \emph{goal-conditioned pruning}:
    $a_t^* = \arg\max_a p(\Phi_t \mid \Phi_{\text{goal}}, a)$,
    which avoids forward simulation by reasoning backward from the goal state.

\textbf{Shared Structure:} 

Both methods transform the exponential search space \( \mathcal{O}(|\mathcal{A}|^T) \) into a tractable inference problem by leveraging structure: 1) In Q-learning: via the Bellman equation; 2) In inverted inference: via latent generative models anchored on goals.
   % \end{itemize}

\textbf{CCUP Perspective:} 

Inverted inference generalizes Q-learning by allowing value propagation to be conditioned on internally specified goals, not just external rewards.

\end{tcolorbox}

\section{Social Symmetry Breaking: Mentalizing and Imitation}
\label{sec:6}

\paragraph{From Symmetric Others to Representational Asymmetry.}

In early evolutionary contexts, all conspecifics were cognitively indistinguishable: other agents were perceived as behaviorally symmetric entities, lacking internal differentiation \cite{baron1997mindblindness}. However, this symmetry presents a fundamental bottleneck for adaptive social interaction: cooperation, competition, teaching, and deception all require the ability to represent other minds as distinct from one’s own.
The emergence of \textbf{mentalizing} \cite{frith2003development}, also known as Theory of Mind (ToM), broke this social symmetry by enabling agents to simulate and infer the internal states of others. This cognitive advance introduced a representational asymmetry between self and other, and further between \emph{known} and \emph{inferred} mental states. The agent can now distinguish between its own beliefs and those it attributes to others, allowing recursive inference: ``I think that you think that I think...'' \cite{premack1978does}.
This was further extended by \textbf{imitation learning} \cite{rajaraman2020toward}, which enabled agents to internalize the behaviors of others by mapping observed actions (ambiguous social context) onto latent generative models (structured content). Through imitation, socially observed behavior is transformed from noisy surface data into reusable policy content.

Under the CCUP framework, the emergence of \textbf{imitation} and \textbf{mentalizing} can also be understood through the principle of \emph{structure-before-specificity}, wherein general representations of agency are first acquired through sensorimotor mirroring \cite{iacoboni2009imitation}, followed by attribution of individual mental states. Formally, we have the following definition.

\begin{definition}[Imitation and Mentalizing in Social Cognition]
    \textbf{Imitation} constructs a shared latent space \( \Phi_{\text{agent}} \) encoding structural priors over actions and goals, which reduces contextual ambiguity: $H(\Psi | \Phi_{\text{agent}}) \ll H(\Psi)$
    by assuming self-other motor symmetry.   
     \textbf{Mentalizing} introduces specificity by distinguishing between known (self) and inferred (other) mental states:
    $\Phi_{\text{self}} \neq \Phi_{\text{other}}, \text{with } \Phi_{\text{other}} \text{ inferred via } p(\Phi_{\text{other}} | \Psi_{\text{self}})$.
\end{definition}
    
The transition from imitation to mentalizing, mirroring the inverted inference in the previous section, reflects a form of cognitive symmetry breaking for social cognition:
$\text{Symmetric Others} \rightarrow \text{Differentiated Minds}$
enabled by the inference pipeline:
$\text{Imitation (Structure)} \Rightarrow \text{Mentalizing (Specificity)}$. It follows that the structure-before-specificity principle boils down to the following proposition.

\begin{proposition}[Imitation-before-Mentalizing in Social Cognition]
Let \( \Phi_{\text{agent}} \) denote a latent structure over social action-goal pairs learned via imitation, and \( \Phi_{\text{other}} \) a specific mental state attributed to another. Under CCUP, general imitation-based structure must precede reliable attribution of mental states:
$H(\Psi | \Phi_{\text{agent}}) \ll H(\Psi | \Phi_{\text{other}}),  \text{unless}  ~\Phi_{\text{other}} \subset \Phi_{\text{agent}}$.
\end{proposition}

\textbf{Remark:}
From the standpoint of CCUP, social cognition entails the transformation of ambiguous social signals \( \Psi \) (facial expressions and gestures) into latent mental representations \( \Phi \) (beliefs and intentions) that are not directly observable \cite{adolphs2018neuroscience}. This process involves minimizing:
$H(\Phi | \Psi) + H(\Psi | \Phi)$ within the context of agent-to-agent inference, where \( H(\Phi | \Psi) \) quantifies the uncertainty in inferring another’s internal state from observed behavior, and \( H(\Psi | \Phi) \) quantifies the ambiguity in reconstructing behavior from inferred states, important for imitation and teaching \cite{meltzoff2005imitation}.
Mentalizing thus implements forward inference, transforming social context into intentional content. Imitation, in turn, reconstructs context from inferred content, completing the inference cycle and enabling the transfer of social knowledge. Such ``inverted inference'' is the bootstrapped consequence of the exploration–exploitation cycle for social cognition. We can observe a similar computational bottleneck arising from exponential representational complexity.

\paragraph{Recursive Inference and Depth of Social Hierarchies.}

Mentalizing supports recursive social inference \cite{gallese2007before}, where the agent models not only the other's mental state, but also the other's model of the agent’s mental state. This recursion introduces a depth hierarchy of social contexts and contents:
$\Psi^{(0)} \rightarrow \Phi^{(0)} \rightarrow \Psi^{(1)} \rightarrow \Phi^{(1)} \rightarrow \cdots$.
Each level in this hierarchy adds to the joint uncertainty but also provides greater flexibility in social prediction, communication, and deception. Under CCUP, such recursion enables the minimization of mutual uncertainty in dyadic or group interactions by aligning nested beliefs and expectations.

\begin{proposition}[Recursive ToM as Hierarchical Inference]
Let \( \Psi^{(k)} \) and \( \Phi^{(k)} \) denote the context and content variables at the \( k \)-th recursive level of ToM, where an agent models the mental state of another agent who is, in turn, modeling the original agent, and so on.
Under the CCUP framework, recursive ToM can be represented as a hierarchy of inference cycles:
$\Psi^{(k)} \rightarrow \Phi^{(k)} \rightarrow \Psi^{(k+1)} \rightarrow \Phi^{(k+1)} \rightarrow \cdots$
Each level \( k \) involves forward inference (disambiguating \( \Phi^{(k)} \) from \( \Psi^{(k)} \), representing the other's belief at level \( k \)) and reverse inference (reconstructing \( \Psi^{(k)} \) from \( \Phi^{(k)} \), simulating the context under which that belief is held).
The total inference uncertainty over \( K \) levels is bounded below by the sum of conditional entropies:
$\sum_{k=0}^{K} \left[ H\left(\Phi^{(k)} | \Psi^{(k)}\right) + H\left(\Psi^{(k)} | \Phi^{(k)}\right) \right] \geq H_{\text{total}} - I_{\text{shared}}$,
where \( H_{\text{total}} \) is the total joint entropy over the social exchange, and \( I_{\text{shared}} \) is the mutual information gained through aligned recursive models.

\end{proposition}

\textbf{Remark:} Deeper ToM reasoning corresponds to deeper context-content cycles, enabling agents to iteratively reduce uncertainty in multi-agent environments by aligning beliefs across nested cognitive levels.
Recursive levels of ToM create a hierarchical inference cycle that deepens social modeling but introduces a computational bottleneck arising from exponential representational complexity. Such a bottleneck can be formalized by the following theorem (its proof sketch is provided in Appendix F).

\begin{theorem}[Recursion Saturation in Theory of Mind]
\label{cor:ToM_saturation}
The recursive embedding of mental models \( \Phi^{(n)} = \text{ToM}^n(\Phi_{\text{other}}) \) leads to exponential growth in representational entropy. Let \( n^* \) be the maximum stable recursion depth such that: $\frac{d H(\Phi^{(n)})}{dn} > 0 \quad \text{but } \quad \frac{d^2 H(\Phi^{(n)})}{dn^2} < 0 \text{ for } n > n^*$.
Then cognitive saturation occurs at \( n^* \), beyond which mentalizing becomes unreliable or collapses into heuristics.
\end{theorem}

\paragraph{From Social Coordination to Cultural Bootstrapping.}

The ability to model and learn from others allows for \emph{social bootstrapping}, where cognitive capacities are enhanced not through isolated trial-and-error, but through shared representations, including pedagogy and imitation. Over evolutionary time, this leads to cultural ratcheting \cite{tennie2009ratcheting}: the progressive accumulation of knowledge across generations.
In CCUP terms, imitation and mentalizing serve to reduce the conditional entropy between one agent’s internal state and another’s observable behavior, enabling distributed inference and coordination:
$H(\Phi_{\text{self}} | \Psi_{\text{other}}) + H(\Psi_{\text{other}} | \Phi_{\text{self}}) \rightarrow \min$.

\begin{proposition}[CCUP Resolution of Recursive ToM via Cultural Bootstrapping]
Let \( \Phi^{(n)} \) denote the $n$-level recursive Theory of Mind (ToM) representation, and let \( \Psi \) denote the observable context. Recursive ToM introduces an entropy bottleneck due to exponential growth in representational complexity:
$\frac{d}{dn} \left[ H(\Psi | \Phi^{(n)}) + H(\Phi^{(n)} | \Psi) \right] > 0  \text{for } n > n^*$.
This bottleneck is resolved via the formation of cyclic, shared structures \( \Phi_{\text{shared}} \), such that:
$H(\Psi | \Phi_{\text{shared}}) + H(\Phi_{\text{shared}} | \Psi) \ll H(\Psi | \Phi^{(n)}) + H(\Phi^{(n)} | \Psi)$.
\end{proposition}

\textbf{Remark:} 
In cultural systems, the latent structure \( \Phi_{\text{shared}} \) becomes externalized and inherited through symbolic communication (language, norms, institutions), allowing distributed inference and the stabilization of social cognition across individuals and generations. The cultural externalization of shared mental structures marks a critical turning point in cognitive evolution: it shifts the burden of deep recursive inference from the individual mind to the collective memory of the group \cite{halbwachs2020collective}.  

\paragraph{Summary.}

Social symmetry breaking occurs when agents begin to model other agents as distinct minds with independent content and context. Through mentalizing and imitation, social interaction is transformed into a recursive inference process, enabling social communication and coordination. These processes reduce uncertainty across agents by aligning ambiguous social cues (context) with structured internal models (content), completing an inference cycle that generalizes beyond the self.

\begin{tcolorbox}[colback=gray!10!white, colframe=blue, title={\textbf{Social Inference: Recursive Theory of Mind and the CCUP Bottleneck}}]

\textbf{The Problem:} \\
Recursive Theory of Mind (ToM) requires modeling others’ beliefs about beliefs, creating a nested inference structure:
$\text{ToM}^{(n)} = \text{``I think that you think that I think...''}$.
As \( n \) increases, the complexity of representing and aligning internal models grows rapidly, leading to a bottleneck in memory, disambiguation, and inference.

\textbf{The Bottleneck:} \\
Each level \( \Phi^{(n)} \) increases representational entropy:
$H(\Phi^{(n)} \mid \Psi) \nearrow \quad \text{as } n \to \infty$,
which eventually saturates, making deeper levels computationally intractable.

\textbf{CCUP Resolution Strategy:}
1) {Cycle Formation}: Turns nested ToM into iterative inference loops that align context and content; 2) {Selective Cloning}: Reuses shared latent templates (roles, prior beliefs) to constrain inference; 3) {Entropy-Based Stopping Rule}: Halts recursion when: $\frac{d}{dn} \left[ H(\Psi \mid \Phi^{(n)}) + H(\Phi^{(n)} \mid \Psi) \right] \geq 0$.
   
\end{tcolorbox}

\section{The Origin of Language: Linguistic Symmetry Breaking}
\label{sec:7}
\paragraph{From Abstract Thought to Structured Expression.} The efficiency of cognitive evolution hinges on the development of a medium capable of transmitting and coordinating ToM's internal representations across agents and time.
Language emerges as a solution to the problem of recursive asymmetry, allowing agents not only to express known mental states but to encode and manipulate inferred or hypothetical ones. In doing so, language breaks the temporal symmetry of thought by introducing {\em structured linear order (syntax) and hierarchical abstraction (grammar)} \cite{halliday1961categories}, enabling the stable transmission of content across varying contextual frames. We now turn to examine how linguistic structures arose as the final and most powerful form of cognitive symmetry breaking.

Before the emergence of language, cognition operated through unstructured or internally simulated representations that were not externally communicable. Mental states, including perceptions, goals, emotions, and memories, lacked standardized formats for interpersonal transfer. Without a mechanism for translating these high-dimensional internal states into shared symbols, thought remained trapped within the individual.
The evolution of language introduced a profound \textbf{linguistic symmetry breaking}, imposing a structured, rule-governed ordering on previously unordered internal representations \cite{fitch2010evolution}. Specifically, it broke the symmetry of temporal and conceptual flow by enforcing linearization (through syntax) and combinatorial constraints (through grammar). What had been simultaneous and non-sequential in thought became serial, parseable, and generalizable across individuals.

\paragraph{Language as Context-Content Alignment.}
Language represents the ultimate form of cognitive symmetry breaking, where the alignment between \emph{context} and \emph{content} reaches its peak. It serves as the bridge between internal thought processes and external communication, allowing individuals to express not only immediate perceptions or emotions but also abstract concepts, future possibilities, and counterfactuals. At its core, language involves the dynamic interplay between contextual cues (what is being referred to or described) and the content (the meaning or intention behind the utterance) \cite{levinson1983pragmatics}. Through grammatical structures such as syntax, morphology, and semantics, language transforms raw sensory inputs and internal thoughts into communicable forms, ensuring that both the \emph{meaning} (content) and the \emph{intended reference} (context) are aligned. This alignment enables agents to share complex mental states, perform reasoning, and engage in recursive thought processes, such as in the formulation of hypothetical situations ("I think that you think that I think..."). As a medium, language facilitates the \emph{resolution of representational asymmetry}, making it possible to shift from self-contained, context-dependent knowledge to universally interpretable, context-structured communication \cite{clark1991grounding}.

Under the CCUP framework, language functions as a tool for minimizing the uncertainty between an agent’s internal state (content \( \Phi \)) and another’s inferred interpretation of it (context \( \Psi \)). In linguistic communication:
1) The speaker maps internal content \( \Phi \) (propositions, intentions) into structured linguistic forms \( \Psi \) (utterances); 2) The listener inverts this mapping, reconstructing \( \Phi \) from \( \Psi \) by minimizing \( H(\Phi | \Psi) \).
This yields a CCUP-style inference cycle across two agents:
$\Phi_{\text{speaker}} \rightarrow \Psi_{\text{utterance}} \rightarrow \Phi_{\text{listener}}$.
Efficient communication requires both agents to jointly minimize:
$H(\Phi_{\text{listener}} | \Psi) + H(\Psi | \Phi_{\text{speaker}})$,
which corresponds to encoding internal ideas into well-structured expressions and recovering them with minimal distortion. Formally, we have

\begin{theorem}[Language as Cross-Agent Completion of the Asymmetric Inference Cycle]
Let \( \Phi^{(A)} \) and \( \Psi^{(A)} \) denote the internal content and context variables of Agent A (the speaker), and \( \Phi^{(B)} \), \( \Psi^{(B)} \) those of Agent B (the listener). Let \( \Psi^{(\text{lang})} \) denote the structured linguistic signal exchanged between them.
Under the CCUP framework, language enables inter-agent inference by establishing a shared latent channel such that
$\Phi^{(A)} \rightarrow \Psi^{(\text{lang})} \rightarrow \Phi^{(B)} \text{and} \quad
\Psi^{(B)} \rightarrow \Phi^{(B)} \rightarrow \Psi^{(A)}$.
This cycle enables:
1) The speaker to compress internal content \( \Phi^{(A)} \) into structured linguistic output \( \Psi^{(\text{lang})} \); 2) The listener to decode \( \Psi^{(\text{lang})} \) into their own internal content \( \Phi^{(B)} \); 3) Mutual refinement of context through recursive alignment, establishing a shared semantic frame.
\label{thm:language}
\end{theorem}

\textbf{Remark:} The proof of the above theorem can be found in Appendix G.
Language completes the CCUP inference hierarchy by enabling distributed context-content cycles across minds. It extends cognitive alignment beyond the individual brain, enabling collective bootstrapping and cumulative culture through structured uncertainty minimization. Language-enabled communication is optimal when the combined uncertainty across agents is minimized \cite{kharitonov2020entropy}:
$H\left( \Phi^{(B)} | \Psi^{(\text{lang})} \right) + H\left( \Psi^{(\text{lang})} |\Phi^{(A)} \right)
+ H\left( \Phi^{(A)} | \Psi^{(B)} \right) + H\left( \Psi^{(B)} | \Phi^{(B)} \right)
\rightarrow \min$. Having established that language enables cross-agent completion of the asymmetric inference cycle, we now turn to how syntax and grammar serve as powerful symmetry breakers, optimizing communication by structuring how meaning is conveyed. Syntax and grammar provide a shared framework that ensures efficient alignment between content and context, allowing for clearer, faster, and more accurate transmission of complex ideas across individuals \cite{gibson2019efficiency}.

\paragraph{Syntax and Grammar as Symmetry Breakers.}

Syntax and grammar impose compositional order over meaning. This structure
linearizes thought across time \cite{fox2005cyclic}, breaking the simultaneity of mental content and disambiguating nested or overlapping ideas via hierarchical parsing.
Moreover, syntax and grammar enable generalization through rule-based recombination (e.g., recursion, agreement, and tense).
In CCUP terms, these constraints reduce ambiguity in both encoding and decoding-
i.e.,  \textbf{speaker:}  $H(\Psi | \Phi) \downarrow$  \text{(structured generation)} and
\textbf{listener:}  $H(\Phi| \Psi) \downarrow$  \text{(structured interpretation)}.
Language does not merely externalize thought but transforms it into a low-entropy, shareable content stream optimized for inter-agent inference \cite{kharitonov2020entropy}. Formally, we have

\begin{proposition}[Syntax and Grammar as Symmetry Breakers in Communication]
Let \( \Phi \) denote the internal representation (content) and \( \Psi \) denote the external context. Without structured communication, \( H(\Psi | \Phi) \) remains high, leading to ambiguity.
{1) Syntax as temporal structure}: 
Syntax orders communication hierarchically (subject-verb-object), breaking temporal symmetry and reducing uncertainty: 
$H(\Psi | \Phi) \ll H(\Psi)$.
{2) Grammar as referential alignment}: 
Grammar resolves referential ambiguity, ensuring clarity in object-action relationships:
$H(\Phi | \Psi) \ll H(\Phi)$.
{3) Recursive grammar as hierarchical structure}: 
Recursive grammar enables nested meanings, breaking contextual symmetry and allowing deeper reasoning:
$\Phi_{\text{self}} \neq \Phi_{\text{other}}, \Phi_{\text{other}} \text{ inferred via recursive ToM}$.
\end{proposition}

\textbf{Remark:} 
Language is hierarchically structured to break symmetry in communication and reduce ambiguity. Syntax imposes temporal order, and grammar clarifies reference, enabling asymmetric mapping between expression and meaning. This structure allows finite rules to generate infinite, interpretable expressions. The Chomsky hierarchy \cite{chomsky1957syntactic} formalizes this progression, with context-free grammar supporting the recursion and embedding essential to human language. From the CCUP perspective, language minimizes uncertainty by aligning content with context across agents. Hierarchical organization thus reflects the cognitive need to transform high-entropy, ambiguous signals into low-entropy, structured meaning for efficient, shared inference.

\paragraph{From Individual Simulation to Collective Bootstrapping.}

Language enables a new form of bootstrapping: not across space (as in steering), time (as in RL), or social inference (as in mentalizing), but across individuals in a group \cite{bennett2023brief}. It turns internal simulations into collective knowledge by enabling distributed content formation across minds.
Through language, internal goals become shared plans $\rightarrow$ Abstract concepts become portable and refinable $\rightarrow$ Cultural memory becomes scalable through symbolic transmission.
This collective bootstrapping mechanism \cite{gentner2010mutual} allows a group to co-construct latent content by aligning noisy individual contexts into common representations. In CCUP terms, language facilitates multi-agent cycle formation, where each member’s content reinforces the context of others.

\begin{corollary}[Cultural Memory as Intergenerational CCUP Compression]
Let \( \{ \Phi_t^{(i)} \} \) denote the internal content representations of individual \( i \) at generation \( t \), and let \( \Psi_t^{(\text{lang})} \) denote the shared linguistic context used to transmit these contents across agents and generations.
Under the CCUP framework, language enables cultural memory by recursively minimizing conditional entropy over time:
$\sum_{t=1}^{T} \left[ H\left(\Phi_t^{(i+1)} \mid \Psi_t^{(\text{lang})} \right) + H\left( \Psi_t^{(\text{lang})} \mid \Phi_t^{(i)} \right) \right] \rightarrow \min$.
This cumulative minimization allows compression of high-dimensional, agent-specific experiences into structured, generalizable linguistic forms \( \Psi_t^{(\text{lang})} \),
reliable reconstruction and elaboration of content \( \Phi_{t+1}^{(j)} \) in future agents or generations, and cultural bootstrapping, whereby each generation inherits a refined prior that reduces their individual inference burden.
\end{corollary}

\textbf{Remark:} Through syntax and grammar, private simulations become communicable, enabling agents to align contexts, share abstractions, and build upon each other’s inferences. This turns solitary learning into cumulative cultural evolution, where knowledge is refined and expanded across minds and generations \cite{kirby2008cumulative}.
Cultural memory emerges as the result of recursively aligned CCUP cycles across individuals and time, with language acting as the stabilizing carrier of low-entropy, high-fidelity content representations in an otherwise noisy and uncertain environment.

\paragraph{Summary.}

Language breaks the symmetry of unstructured thought by imposing an ordered, combinatorial form. It transforms ambiguous internal content into compressible, generalizable, and recoverable structures that minimize uncertainty in communication. Under CCUP, this transformation enables cross-agent alignment via recursive inference cycles, allowing cognition to scale from individual simulation to cultural computation. Syntax and grammar are not arbitrary constraints; they are evolution’s solution to stabilizing meaning in an uncertain world.

\begin{tcolorbox}[colback=gray!10!white, colframe=blue, title={\textbf{Chomsky Hierarchy: Linguistic Symmetry Breaking}}]

The Chomsky hierarchy classifies formal languages by increasing generative power:

\begin{itemize}
    \item \textbf{Type 3 (Regular Languages)}: Encodes linear sequences; represents symmetry breaking in \emph{temporal ordering}, where one symbol follows another. Suitable for simple reactive behaviors.
    
    \item \textbf{Type 2 (Context-Free Languages)}: Introduces nested structure; breaks symmetry by allowing \emph{hierarchical grouping} (e.g., subject-verb-object). Enables parsing of constituent structure in phrases.
    
    \item \textbf{Type 1 (Context-Sensitive Languages)}: Breaks \emph{contextual uniformity} by allowing rules to depend on neighboring symbols—facilitating agreement and disambiguation in complex utterances.
    
    \item \textbf{Type 0 (Recursively Enumerable)}: Represents the most general form of linguistic generation, capable of simulating arbitrary computation. Breaks the boundary between language and logic.
\end{itemize}

\vspace{0.5em}

\textbf{From a CCUP Perspective:} \\
Each transition up the Chomsky hierarchy corresponds to a deeper level of symmetry breaking:
%\begin{align*}
\text{Linear Time} $\rightarrow$ \text{Nested Syntax};
\text{Uniform Context} $\rightarrow$ \text{Conditional Structure};
\text{Finite Expression} $\rightarrow$ \text{Unbounded Computation}.
%\end{align*}
This reflects the gradual increase in context-content alignment complexity required for expressing internal states, coordinating shared knowledge, and communicating recursive beliefs. Language thus serves as the final bootstrapping mechanism for resolving the uncertainty bottlenecks of social and cultural inference.

\end{tcolorbox}

\section{Conclusion: Broken Symmetry in Cognition}

The principle of \emph{broken symmetry} plays a fundamental role in shaping the cognitive architecture of biological systems. In early evolutionary contexts, organisms faced a world where sensory input was often symmetric, with little distinction between self and other, or between immediate and future states. Over time, however, evolutionary pressures necessitated the emergence of \textbf{representational asymmetries}, from the temporal asymmetry of predictive inference to the social asymmetry of theory of mind (ToM). These asymmetries break the symmetry of raw sensory input and organize it into structured, actionable representations.
At the heart of these cognitive breakthroughs lies the idea of \textbf{context-content alignment}, which resolves the information bottleneck in high-entropy environments. The transition from symmetric to asymmetric representations begins with spatial symmetry breaking, where simple sensory input becomes organized into structured representations like grid and place cells. These spatial anchors provide the scaffold for temporal symmetry breaking, enabling predictive dynamics (e.g., reinforcement learning), and eventually paving the way for higher-order cognitive functions like mentalizing and language.

The emergence of language marks the pinnacle of cognitive symmetry breaking, where syntax and grammar enable agents to efficiently communicate complex, recursive mental states. Through the lens of the \textbf{Context-Content Uncertainty Principle (CCUP)}, language acts as a bridge, transforming internal representations into shared, communicable forms that resolve the ambiguity between context (the external world) and content (internal thought). The recursive structure of language, including its syntactic and grammatical rules, allows for the efficient transfer of knowledge, ensuring that social agents can communicate, cooperate, and engage in complex reasoning.
In summary, broken symmetry in cognition is not just a feature of individual mental processes, but a \textbf{dynamic interaction} between self, other, and shared environment. From early spatial navigation to the development of theory of mind and language, symmetry breaking is the key to resolving uncertainty, enabling complex cognition, and facilitating adaptive behavior in social contexts. The study of this principle offers profound insights into the evolution of intelligence, as well as the computational strategies that support it across species and systems.

\bibliographystyle{ieeetr}
\bibliography{ref}

%\newpage
\appendix

\renewcommand{\theequation}{A\arabic{equation}}
\renewcommand{\thefigure}{A\arabic{figure}}
\renewcommand{\thetable}{A\arabic{table}}
\setcounter{equation}{0}    
\setcounter{figure}{0}    
\setcounter{table}{0}    

% \onecolumn
\section{Proof of Theorem 1}
\label{appendix:A}

\begin{proof}
We aim to show that under the Context-Content Uncertainty Principle (CCUP), minimizing \( H(\Psi \mid \Phi) \) to construct a generalizable structure \( \Phi \) prior to minimizing \( H(\Phi \mid \Psi) \) for content inference leads to lower inference error and more stable representations.

\textbf{Step 1: Entropic Asymmetry under CCUP.} \\
By CCUP, context variables \( \Psi \) are high-entropy and ambiguous, while content variables \( \Phi \) are lower-entropy and selectively encoded. That is,
$H(\Psi) \gg H(\Phi)$,
which implies:
$H(\Phi \mid \Psi) > H(\Psi \mid \Phi)$.
From the definition of mutual information:
$I(\Psi; \Phi) = H(\Psi) - H(\Psi \mid \Phi) = H(\Phi) - H(\Phi \mid \Psi)$,
it follows that minimizing \( H(\Psi \mid \Phi) \) yields greater gains in mutual information when \( \Phi \) is still underdetermined.

\textbf{Step 2: Generalizable Structure Minimizes Cross-Context Variance.} \\
Let \( \Phi^\ast \) be the representation that minimizes expected context entropy across observations:
$\Phi^\ast = \arg\min_{\Phi} \mathbb{E}_{i=1}^N \left[ H(\Psi_i \mid \Phi) \right]$.
This optimization extracts invariant latent structure across variable contexts. For any new context \( \Psi_j \), this structural prior ensures that:
$\mathbb{E}_j \left[ H(\Psi_j \mid \Phi^\ast) \right] < \mathbb{E}_j \left[ H(\Psi_j \mid \Phi_{\text{early}}) \right]$,
where \( \Phi_{\text{early}} \) is a context-specific representation inferred without prior structure, which is more prone to overfitting and aliasing.

\textbf{Step 3: Specificity after Structure Minimizes Inference Error.} \\
Let the expected inference loss be:
$\mathcal{L} = \mathbb{E}_j \left[ H(\Phi \mid \Psi_j) \right]$.
By the chain rule of entropy:
$H(\Phi \mid \Psi_j) = H(\Phi \mid \Psi_j, \Phi^\ast) + I(\Phi ; \Phi^\ast \mid \Psi_j)$.
Since \( \Phi^\ast \) is a learned structural prior that compresses variability across contexts, it reduces the conditional entropy of \( \Phi \) given \( \Psi_j \):
$H(\Phi \mid \Psi_j, \Phi^\ast) < H(\Phi \mid \Psi_j)$.
Therefore, conditioning inference on prior structure strictly improves specificity:
$\mathcal{L}_{\text{with structure}} < \mathcal{L}_{\text{without structure}}$.

\textbf{Conclusion.} \\
Under CCUP, first minimizing \( H(\Psi \mid \Phi) \) enables generalization through structure formation, and only then minimizing \( H(\Phi \mid \Psi) \) enables robust, low-error specificity. Hence, structure must precede specificity for effective inference and memory.
\end{proof}

\section{Proof of Theorem 2}
\label{appendix:B}

\begin{proof}[Proof of Corollary~\ref{cor:delay_enables_generalization}]
Let \( \mathcal{D}_{\text{train}} = \{ \Psi_i \}_{i=1}^N \) be a diverse set of contexts. Let the system construct a structural representation \( \Phi^\ast \) by minimizing expected conditional entropy:
$\Phi^\ast = \arg\min_{\Phi} \mathbb{E}_i \left[ H(\Psi_i \mid \Phi) \right]$,
which identifies latent structure shared across multiple contexts.

Now consider a new context \( \Psi_j \notin \mathcal{D}_{\text{train}} \). Two strategies are available:

\begin{itemize}
    \item \textbf{Delayed Binding:} Use \( \Phi^\ast \) to perform inference by minimizing \( H(\Phi^\ast \mid \Psi_j) \).
    \item \textbf{Premature Binding:} Infer \( \Phi_{\text{early}} \sim p(\Phi \mid \Psi_j) \) directly without prior structure.
\end{itemize}

We compare the expected inference uncertainty in both cases:
$H(\Phi_{\text{early}} \mid \Psi_j) \quad \text{vs.} \quad H(\Phi^\ast \mid \Psi_j)$.

Because \( \Phi^\ast \) is learned via structural compression across contexts, it captures shared latent invariants and reduces the effective hypothesis space. By contrast, \( \Phi_{\text{early}} \) reflects an under-regularized inference from ambiguous, high-entropy context \( \Psi_j \). Therefore, by the information-theoretic principle of \textit{bias-variance tradeoff}:
$\mathbb{E}_j \left[ H(\Phi_{\text{early}} \mid \Psi_j) \right] > \mathbb{E}_j \left[ H(\Phi^\ast \mid \Psi_j) \right]$.

Further, if the learning system attempts to bind content before structure is formed, the resulting representations are more sensitive to context noise and less transferable to novel inputs. This leads to:

\begin{itemize}
    \item \textbf{Increased variance:} Context-sensitive representations vary across similar inputs.
    \item \textbf{Reduced generalization:} Learned representations do not support inference on unseen or aliased contexts.
\end{itemize}

Therefore, early binding without structure leads to higher conditional entropy and inference error:
\[
\mathbb{E}_j \left[ H(\Phi \mid \Psi_j, \Phi_{\text{early}}) \right] > \mathbb{E}_j \left[ H(\Phi \mid \Psi_j, \Phi^\ast) \right].
\]

\textbf{Conclusion.} Delaying specificity until after structure formation leads to more robust generalization, lower entropy, and reduced inference error across contexts. Hence, under CCUP, structural delay is not a weakness but a computational necessity for effective learning.
\end{proof}

\section{Proof of Theorem 3}
\label{appendix:C}

\begin{proof}
We prove the theorem in three logical steps, corresponding to its three claims.

\textbf{Step 1: Spatial symmetry is broken by internal structuring.} \\
Let \( \Psi \) denote the raw, high-entropy sensorimotor context of an organism, and let \( \Phi_{\text{space}} \) denote its latent spatial representation. In an isotropic environment, \( \Psi \) is undirected and ambiguous:
$H(\Psi) \gg H(\Phi_{\text{space}})$,
provided that \( \Phi_{\text{space}} \) is formed via structure-learning (e.g., grid cells, place fields). The process of spatial symmetry breaking thus corresponds to constructing a compressed and low-entropy representation:
$H(\Phi_{\text{space}}) \ll H(\Psi)$.
This spatial compression reduces aliasing and disambiguates direction, establishing consistent frames of reference such as "forward" or "left".

\textbf{Step 2: Temporal inference bootstraps on spatial structure.} \\
Let \( \Phi_{\text{time}} \) be the content variable representing temporally extended expectations, such as value or future state. In reinforcement learning, predicting future rewards or outcomes requires modeling conditional transitions between states. If the underlying space is structured by \( \Phi_{\text{space}} \), then temporal inference can be conditioned on that structure:
$H(\Phi_{\text{time}} \mid \Phi_{\text{space}}) \ll H(\Phi_{\text{time}} \mid \Psi)$,
since \( \Phi_{\text{space}} \) reduces uncertainty about state transitions. Without such a scaffold, inference over time becomes ill-posed due to aliasing in raw context \( \Psi \). Therefore, spatial anchoring provides the necessary prior for effective temporal compression.

\textbf{Step 3: Temporal asymmetry depends on spatial anchoring.} \\
Directional asymmetry in time (e.g., favoring future over past, discounting) requires a stable internal model of change. Such a model must operate over well-defined states. If \( \Phi_{\text{space}} \) is not yet structured (i.e., the organism lacks an internal map), then any attempt to assign future expectations \( \Phi_{\text{time}} \) to raw input \( \Psi \) suffers from overfitting or representational interference:
$\text{If } \Phi_{\text{space}} \text{ is unstructured, then } H(\Phi_{\text{time}} \mid \Psi) \text{ is high and unstable.}$
This leads to the failure of stable value estimation or goal prediction. Hence, spatial structuring is a necessary precondition for breaking temporal symmetry through reinforcement learning.

\textbf{Conclusion.} \\
We have shown that (1) structured spatial representations reduce the entropy of raw context, (2) temporal inference conditioned on spatial structure is more stable and generalizable, and (3) temporal asymmetry cannot emerge without prior spatial anchoring. Thus, spatial symmetry breaking bootstraps the emergence of directional inference in time.
\end{proof}

\section{Proof of Theorem 4}
\label{appendix:D}

\begin{proof}
(1) Pattern recognition identifies stable features (e.g., objects, locations) from noisy, aliased sensory input \( \Psi \), compressing the high-dimensional input into structured spatial representations \( \Phi_{\text{space}} \). This breaks spatial symmetry and reduces entropy:
$H(\Phi_{\text{space}}) \ll H(\Psi)$;

(2) Temporal inference, governed by RL, relies on predicting future states and rewards. With spatial structure established, RL can condition temporal predictions on the stabilized latent space:
$H(\Phi_{\text{time}} \mid \Phi_{\text{space}}) \ll H(\Phi_{\text{time}} \mid \Psi)$.
This reflects the CCUP claim that spatially grounded context disambiguates temporally unfolding content.

(3) The CCUP framework defines inference as a bidirectional cycle: bottom-up recognition \( p(\Phi \mid \Psi) \) and top-down prediction \( p(\Psi \mid \Phi) \). As RL and perception co-adapt, the joint cycle minimizes uncertainty:
$H(\Phi \mid \Psi) + H(\Psi \mid \Phi) \to \min$.
This convergence resolves the information bottleneck and constructs a world model in which temporal asymmetry (e.g., preference for future rewards) is made possible by prior spatial anchoring.

\end{proof}

\section{Proof of Theorem 5}
\label{appendix:E}

\begin{proof}
Let \( \Phi_{0:T} = (\Phi_0, \Phi_1, \ldots, \Phi_T) \) be a trajectory in latent space. In forward simulation, the agent begins at \( \Phi_0 \) and samples actions to generate state transitions. The number of possible trajectories grows exponentially with horizon \( T \), leading to a combinatorial explosion in \( \dim(\Phi)^T \).

Let us define:
$\mathcal{S}_{\text{forward}} = \{ \Phi_{0:T} \mid \Phi_0 \text{ fixed} \}, \quad
\mathcal{S}_{\text{inv}} = \{ \Phi_{0:T} \mid \Phi_T = \Phi_{\text{goal}} \text{ fixed} \}$.

\textbf{Step 1 (Search space complexity):}  
Because goal conditioning restricts inference to only those paths that lead to \( \Phi_{\text{goal}} \), we have:
$|\mathcal{S}_{\text{inv}}| \ll |\mathcal{S}_{\text{forward}}|, \quad \Rightarrow \dim(\mathcal{M}_{\text{goal}}) \ll \dim(\Phi)^T$

\textbf{Step 2 (Entropy reduction):}  
Since the support of the conditional distribution \( p(\Phi_{0:T} \mid \Phi_{\text{goal}}) \) is strictly contained within the support of \( p(\Phi_{0:T} \mid \Phi_0) \), and goal-conditioned inference excludes many irrelevant trajectories, it follows that:
$H(\Phi_{0:T} \mid \Phi_{\text{goal}}) < H(\Phi_{0:T} \mid \Phi_0)$.

\textbf{Step 3 (CCUP application):}  
Under the CCUP framework, goal \( \Phi_{\text{goal}} \) functions as a low-entropy content anchor. Conditioning on it allows the agent to select a minimal uncertainty trajectory that satisfies the desired constraint, minimizing:
$H(\Phi_{0:T} \mid \Phi_{\text{goal}}) + H(\Phi_{\text{goal}} \mid \Phi_{0:T}) \to \min$.

This bidirectional reduction defines a cycle-consistent inference loop where both perception and planning are anchored on structured content.

\textbf{Step 4 (Conclusion):}  
Thus, inverted inference improves the tractability and precision of goal simulation by:
1) Restricting inference to goal-aligned manifolds; 2)
Minimizing conditional entropy of sampled trajectories; 3)
 Avoiding exhaustive forward expansion from arbitrary initial states.
%\end{itemize}

This completes the proof.
\end{proof}

\section{Proof of Theorem \ref{cor:ToM_saturation}}
\label{appendix:F}

\begin{proof}
Let \( \{(\Psi^{(k)}, \Phi^{(k)})\}_{k=0}^{K} \) represent a hierarchy of context-content inference cycles across \( K \) recursive levels of Theory of Mind, as defined in Proposition 9.

Recursive ToM inference saturates at depth \( K^* \) if, for all \( k \geq K^* \), the marginal gain in mutual information becomes negligible:
$I\left(\Psi^{(k+1)} ; \Phi^{(k+1)}\right) - I\left(\Psi^{(k)} ; \Phi^{(k)}\right) \leq \epsilon, \quad \text{for some } \epsilon > 0$.
Equivalently, the incremental uncertainty reduction flattens:
$\Delta_k := \left[ H\left(\Phi^{(k)} \mid \Psi^{(k)}\right) + H\left(\Psi^{(k)} \mid \Phi^{(k)}\right) \right] - 
\left[ H\left(\Phi^{(k+1)} \mid \Psi^{(k+1)}\right) + H\left(\Psi^{(k+1)} \mid \Phi^{(k+1)}\right) \right] \leq \delta$
for small \( \delta > 0 \). This saturation implies that beyond depth \( K^* \), further recursion yields diminishing cognitive utility, either due to bounded rationality, communication limits, or shared priors between agents.

Thus, recursive social inference is effectively bounded in depth, and CCUP provides a principled information-theoretic criterion for when additional levels of ToM cease to be informative.
\end{proof}

\section{Proof of Theorem \ref{thm:language}}
\label{appendix:G}

\begin{proof}
We begin with the CCUP framework, where cognition operates by cyclically minimizing the conditional entropies \( H(\Phi \mid \Psi) \) and \( H(\Psi \mid \Phi) \). In the intra-agent setting, this inference loop aligns latent context with observed or imagined content.

In the inter-agent setting, language serves as the structured communication channel \( \Psi^{(\text{lang})} \), permitting the inference cycle to extend across agents.

\textbf{Step 1: Compression (Speaker Encoding)} \\
The speaker performs a mapping \( f_A: \Phi^{(A)} \mapsto \Psi^{(\text{lang})} \), compressing internal content into a signal optimized for interpretability. This satisfies:
$H(\Psi^{(\text{lang})} \mid \Phi^{(A)}) \ll H(\Psi^{(\text{lang})})$,
i.e., the linguistic signal carries information about the speaker's internal state.

\textbf{Step 2: Reconstruction (Listener Decoding)} \\
The listener maps \( \Psi^{(\text{lang})} \) to their own content estimate \( \Phi^{(B)} \) via \( f_B^{-1}: \Psi^{(\text{lang})} \mapsto \Phi^{(B)} \). If successful,
$H(\Phi^{(B)} \mid \Psi^{(\text{lang})}) \ll H(\Phi^{(B)})$,
showing the listener can infer content from the speaker’s signal.

\textbf{Step 3: Contextual Alignment} \\
The listener’s inference updates \( \Psi^{(B)} \), and through recursive modeling of \( \Psi^{(A)} \), further refines the mapping:
$\Psi^{(B)} \rightarrow \Phi^{(B)} \rightarrow \Psi^{(A)}$.
This recursive structure enables dynamic re-alignment of context between agents, minimizing mismatch across communicative turns:
$\min_{\Psi^{(\text{shared})}} \left[ H(\Phi^{(A)} \mid \Psi^{(\text{shared})}) + H(\Phi^{(B)} \mid \Psi^{(\text{shared})}) \right]$.

\textbf{Conclusion:} \\
By forming a shared inference cycle through \( \Psi^{(\text{lang})} \), language enables distributed minimization of uncertainty, effectively closing the CCUP inference loop across agents. This transforms asymmetric, private mental states into coordinated, recursively aligned knowledge—completing the inference cycle at the inter-agent level.
\end{proof}

\end{document}